\documentclass[twocolumn,aps,prd,groupedaddress]{revtex4}
\usepackage{amssymb,amsmath}
\usepackage{epsfig}
\usepackage{epstopdf}
\usepackage{tikz,tikz-3dplot}

\begin{document}
\title{Full analytic expression of overlap reduction function for gravitational wave background with pulsar timing arrays}

\author{Yu Hu}
\author{Pan-Pan Wang}
\author{Yu-Jie Tan}

\author{Cheng-Gang Shao}
 \email{cgshao@hust.edu.cn}
\affiliation{MOE Key Laboratory of Fundamental Physical Quantities Measurement, Hubei Key Laboratory of Gravitation and Quantum Physics, PGMF, and School of Physics, Huazhong University of Science and Technology - 430074, Wuhan, Hubei, China}

\begin{abstract}
    Pulsar timing array (PTA) is expected to detect gravitational wave background (GWB) in the nanohertz band within the next decade.
    This provides an opportunity to test the gravity theory and cosmology.
    A typical data analysis method to detect GWB is cross-correlation analysis.
    The overlap reduction function (ORF) plays an important role in the correlation data analysis of GWB.
    The present approach to dealing with the intricate integration in ORF is to use short-wave approximation to drop out the tricky terms.
    In this paper, we provide the full analytic expression of the ORF for PTA without any approximation for all possible polarizations allowed by modifications of general relativity.
    Compared with the numerical simulation and short-wave approximation, our results are more efficient and widely applicable.
    Especially for the scalar-longitudinal mode where the short-wave approximation is not available, our analytical expression is particularly significant. 
\end{abstract}

\maketitle

\section{Introduction\label{sec1}}
A lot of gravitational waves (GWs) induced by compact binary mergers \cite{2016prl_GW150914,2017prl_GW170817,2021apjl_NSBH} have been detected by LIGO and Virgo, which opens a new window for astronomical observations \cite{2019prx_GWTC-1,2021prx_GWTC-2}.
The GWs can not only provide information about the compact binary, but also provide an opportunity to test the gravity theory and cosmology \cite{2019prd_test_GR_GWTC-1}.
These abundant GW signals imply that there are many weak GWs that the detector cannot discern.
These weak GWs are indeed recorded by the detecter, but they are just masked by noise.
The combined weak signal from the population of binary black holes is an example of gravitational wave background (GWB).
In addition, there are many other sources that can generate GWB, such as cosmological phase transitions \cite{1986mnras_GWB_cosmol_pt,2021prd_GWB_QCD_pt}, primordial gravitational waves \cite{1992pr_cosmol_perturbation,1997prd_GWB_inflation}, cosmic strings \cite{2005prd_GWB_cosmic_strings,2007prl_GWB_cosmic_strings}, etc.
The detection of the GWB can provide us with information on the astronomical distribution and the early evolution of the universe, which is of great significance.

At present, the ground-based laser interferometers cannot detect the GWB, and give an upper limits of dimensional energy density $\Omega_{\text{GW}}\leq5.8\times 10^{-9}$ at Hz band \cite{2021prd_upper_limits_GWB_LIGO_O3}.
The future space-based interferometers such as LISA \cite{2017arX_LISA} and TianQin \cite{2016cqg_TQ} probe GWB at the mHz band.
Pulsar timing arrays (PTA) probe GWB at much lower frequency of nHz and appears to be on the verge of detecting GWB now \cite{2020apjl_GWB_NANOGrav,2021mnras_GWB_EPTA,2021apjl_GWB_PPTA,2022mnras_GWB_IPTA}.
The principle of the PTA is simple: the radio waves emitted by the rotating neutron star sweep through the earth at regular interval, so the time interval between the pulses arriving at the detectors on Earth is supposed to be constant.
However, the GWs on the path cause the deviation from a regular pulse.

Cross-correlation is a typical method to detect the GWB.
In a single detector, the GWB would be masked by noise, making it impossible to discern it from noise.
However, the GWB signals in different detectors are correlated, and the noise is not, so the GWB signals can be extracted by correlating the outputs of two different detectors.
To obtain the correlation signal, we need to calculate the correlation coefficient, which is the ORF.
The ORF of PTA is the geometric factor of the correlation signal, which only depends on the relative orientation of the two pulsars and their distance to the Earth.

In general, one has to evaluate the integral in ORF numerically, due to the non-trivial exponential terms.
With the short-wave approximation, the integral can be done analytically by dropping out the exponential terms.
The results ORF for tensor mode is the Hellings and Downs curve \cite{1983apj_HD_curve}.
The approximate ORF for vector mode and scalar-breathing mode can be found in a similar way \cite{2008apj_Lee_ORF_nonGR}.
In the limit that the angle between two pulsars is close to 0, the short-wave expression of the ORF for the vector mode diverges, which means that the pulsar terms dropped out should be included to handle that case.
Furthermore, for the scalar-longitudinal ORF, there is no analytic expression with the short-wave approximation except for some special cases \cite{2012prd_ORF_nonGR,2015prd_PTA_GWB_nonGR}.
Because dropping out the exponential terms and integrating the remaining part will yield a divergent result.
Recently, an analytic series expansion of ORF for all six polarizations was obtained without employing the short-wave approximation\cite{2021prd_ORF_PTA_series_GR,2022prd_ORF_PTA_series_nonGR}.
However, the series expansion is not satisfactory, because it is relatively complex and takes a long time to calculate.
We find that the full analytic expressions of ORF for all six polarizations without employing the short-wave approximation can be obtained using a special integration method, which is used to calculate the response function of space-borne gravitational wave detectors \cite{2019prd_analytic_response_function_TDI,2021prd_sensitivity_TDI,2021prd_sensitivity_TDI_nonGR}.

The paper is organized as follows.
In Sec.\ref{sec2}, we briefly describe the correlations of PTA signal and outline the integration of the ORF.
In Sec.\ref{sec3}, we derive the analytic expansions for ORF for all polarizations and analyze their performance from different perspectives.
Finally, in Sec. \ref{sec4}, we present a brief discussion.

\section{CORRELATIONS OF PTA SIGNALS\label{sec2}}

\subsection{GWB statistic}
The metric perturbations corresponding to GWB can be expressed as 
\begin{equation}\label{h_ab_t}
    \begin{aligned}
    h_{ab}(t,\vec{x})=\int_{-\infty}^{\infty}df \int d^2\Omega_{\hat{n}} &
    \sum_A h_{A}(f,\hat{n}) \\
    &\times e^{A}_{ab}(\hat{n}) e^{i2\pi f(t+\hat{n} \cdot \vec{x}/c)},
    \end{aligned}
\end{equation}
where $e^{A}_{ab}(\hat{n})$ is the spin-2 polarization tensors.
$A=\{+, \times, X, Y, B, L\}$ represent different polarization mode, where ${+,\times}$ represent tensor mode predicted by general relativity, ${X,Y}$ and ${B,L}$ represent vector mode and scalar allowed by the general metric theory of gravity.
Explicitly,
\begin{equation}\label{e_ab_n}
    \begin{aligned}
    e^{+}_{ab}(\hat{n})=&\hat{\theta}_a \hat{\theta}_b - \hat{\phi}_a \hat{\phi}_b , \quad
    e^{\times}_{ab}(\hat{n})=\hat{\theta}_a \hat{\phi}_b + \hat{\phi}_a \hat{\theta}_b ,\\
    e^X_{ab}(\hat{n})=&\hat{\theta}_a\hat{n}_b+\hat{n}_a\hat{\theta}_b ,\quad
    e^Y_{ab}(\hat{n})=\hat{\phi}_a\hat{n}_b+\hat{n}_a\hat{\phi}_b ,\\
    e^B_{ab}(\hat{n})=&\hat{\theta}_a\hat{\theta}_b+\hat{\phi}_a\hat{\phi}_b ,\quad
    e^L_{ab}(\hat{n})=\sqrt{2}\hat{n}_a\hat{n}_b ,
\end{aligned}
\end{equation}
where
\begin{equation}\label{n}
    \begin{aligned}
        \hat{n}     &=(\sin\theta \cos\phi,\sin\theta \sin\phi,\cos\theta),\\
        \hat{\theta}&=(\cos\theta \cos\phi,\cos\theta \sin\phi,-\sin\theta) ,\\
        \hat{\phi}  &=(\sin\phi,\cos\phi,0) .
    \end{aligned}
\end{equation}

For the isotropic, unpolarized and stationary GWB, the quadratic expected values satisfy that
\begin{equation}\label{S_h}
    \left\langle h_{A}(f, \hat{n}) h_{A^{\prime}}^{*}\left(f^{\prime}, \hat{n}^{\prime}\right)\right\rangle
    =\frac{1}{8 \pi} S^A_{h}(f) \delta\left(f-f^{\prime}\right) \delta_{A A^{\prime}} \delta^{2}\left(\hat{n}, \hat{n}^{\prime}\right) ,
\end{equation}
where $S^A_{h}(f)$ can be regarded as the component corresponding to the $A$ polarization of one-sided gravitational-wave strain power spectral density function.
For tensor mode and vector mode, it can be considered that $S^{+}_{h}=S^{\times}_{h}=S^{T}_{h}/2$, $S^{X}_{h}=S^{Y}_{h}=S^{V}_{h}/2$.
However, the two scalar modes should be considered as two independent polarization modes.
The strain power spectral density is simply related to the fractional energy density,
\begin{equation}\label{S_h_Omega}
    S_h(f)=\frac{3H_0^2}{2\pi^2}\frac{\Omega_{\mbox{gw}}(f)}{f^3}.
\end{equation}
The fractional energy density is defined as
\begin{equation}\label{Omega}
    \Omega_{\mbox{gw}}(f)=\frac{1}{\rho_{c}} \frac{d \rho_{\mbox{gw}}}{d \ln f} 
\end{equation}
where $\rho_{c} \equiv 3 c^{2} H_{0}^{2} / 8 \pi G $ is the critical energy density required for closed universe, and $H_0$, $G$ are the Hubble constant and gravitational constant respectively.

\subsection{Correlation signal}
The metric perturbation is weak, so the single pulsar timing signal can be written as 
\begin{equation}\label{h_t}
    h(t)=\int_{-\infty}^{\infty} d f \int d^{2} \Omega_{\hat{n}} R^{a b}(f, \hat{n}) h_{a b}(f, \hat{n}) e^{i 2 \pi f t}.
\end{equation}
Here the response function for Doppler frequency measurements is \cite{2017Review_detection_GWB}
\begin{equation}\label{R_ab}
    R^{ab}(f,\hat{n})=\frac{1}{2} \frac{u^a u^b}{1+\hat{n}\cdot\hat{u}} \left(1-e^{-\frac{i2\pi fL}{c}(1+\hat{n}\cdot\hat{u})}\right) e^{i2\pi f\hat{n}\cdot\vec{r}_2/c} ,
\end{equation}
where $\hat{u}$ is the unit vector of the pulsar pointing to the earth, $L$ is the distance from the pulsar to the earth, and $\vec{r}_2$ is the position vector of the earth.
In the frequency domain, the signal is written in terms of polarized basis as
\begin{equation}\label{h_f}
    \tilde{h}(f)=\int d^2\Omega_{\hat{n}} \sum_A R^A(f,\hat{n})h_A(f,\hat{n}),
\end{equation}
where $R^A(f,\hat{n})= R^{ab}(f,\hat{n})e^A_{ab}(\hat{n})$.

The cross-correlation of two pulsars signal $I$ and $J$ is
\begin{equation}\label{h_I_h_J}
    \left\langle \tilde{h}_I(f)\tilde{h}_J^*(f^\prime)\right\rangle 
    =\frac{1}{2}\delta(f-f^\prime) \sum_A\Gamma^A_{IJ}(f)S^A_h(f),
\end{equation}
where $A=\{ T,V,B,L \}$.
The ORF for tensor mode, vector mode, scalar-breathing mode and scalar-longitudinal is
\begin{equation}
    \begin{aligned}
    \Gamma_{IJ}^T(f)&=\frac{1}{8\pi}\int d^2\Omega_{\hat{n}}\sum_{A=+,\times} R^A_I(f,\hat{n})R^{A*}_J(f,\hat{n}) ,\\
    \Gamma_{IJ}^V(f)&=\frac{1}{8\pi}\int d^2\Omega_{\hat{n}}\sum_{A=X,Y} R^A_I(f,\hat{n})R^{A*}_J(f,\hat{n}) ,\\
    \Gamma_{IJ}^B(f)&=\frac{1}{4\pi}\int d^2\Omega_{\hat{n}} R^B_I(f,\hat{n})R^{B*}_J(f,\hat{n}) ,\\
    \Gamma_{IJ}^L(f)&=\frac{1}{4\pi}\int d^2\Omega_{\hat{n}} R^L_I(f,\hat{n})R^{L*}_J(f,\hat{n}) .
\end{aligned}
\end{equation}

\section{Overlap reduction function\label{sec3}}
\begin{figure}[!t]
    \centering
    \includegraphics[width=0.3\textwidth]{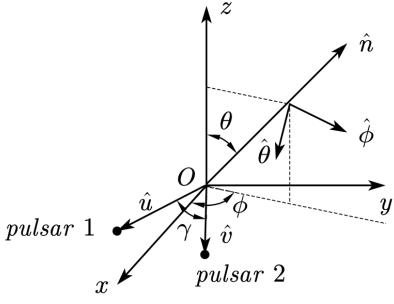}
    \caption{Our convention for the coordinate. }\label{fig}
\end{figure} 
The ORF only depends on the relative orientation of the two pulsars and their distance to the Earth.
Therefore, we can choose a suitable coordinate system to simplify the integral.
We choose the Earth as the coordinate origin, so that $\vec{r}_2=0$.
And the direction vectors of the two pulsars are $\hat{u}=(\cos\frac{\gamma}{2},\sin\frac{\gamma}{2},0)$,
$\hat{v}=(\cos\frac{\gamma}{2},-\sin\frac{\gamma}{2},0 )$, where $\gamma$ is the angle separation between two pulsars.
For the convenience of calculation, we define that
\begin{equation}
    \begin{aligned}
    x=&\hat{n}\cdot\hat{u}=\sin\theta\cos\tilde{\phi} ,\\
    y=&\hat{n}\cdot\hat{v}=\sin\theta\cos\underset{\sim}{\phi},
    \end{aligned}
\end{equation}
where $\underset{\sim}{\phi} = \phi-\frac{\gamma}{2} $, $\tilde{\phi} = \phi+\frac{\gamma}{2} $.
\subsection{Tensor mode}
The ORF of the tensor mode is 
\begin{equation}\label{ORF_T}
    \begin{aligned}
    \Gamma&^T_{IJ}(f)=\Gamma^T_{IJ}(\beta_u,\beta_v,\gamma) \\
    &=\frac{1}{32\pi}\int d^2\Omega_{\hat{n}}
    \frac{(1-e^{-i\beta_u (1+\sin\theta\cos\underset{\sim}{\phi})})}
    {(1+\sin\theta\cos\underset{\sim}{\phi}) }\\
    &\times \frac{(1-e^{i\beta_v (1+\sin\theta\cos\tilde{\phi})})} 
    {(1+\sin\theta\cos\tilde{\phi}) } \\
    &\times ((1-\sin^2\theta\cos^2\underset{\sim}{\phi})(1-\sin^2\theta\cos^2\tilde{\phi})
    -2\cos^2\theta\sin^2\gamma) ,
    \end{aligned}
\end{equation}
where $\beta_u=2\pi f L_u/c$, $\beta_v=2\pi f L_v/c$, and $L_u, L_v$ are the distances from the two pulsars to the Earth, respectively.
Change the integral variables $\{\theta,\phi\}$ by $\{x,y\}$, so that the integral region is transformed from the unit sphere to the elliptical area on the $xy$ plane.
Then,
\begin{equation}\label{ORF_T_xy}
    \begin{aligned}
    \Gamma&^T_{IJ}(f)=\frac{1}{16\pi}\int_{-1}^1dx\int_{y_l}^{y_u}dy
    (1-e^{-i\beta_u (1+x)})\\
    &\times (1-e^{i\beta_v (1+y)}) 
    \left( \frac{(1-x)(1-y)}{\eta(x,y)}-\frac{2\eta(x,y)}{(1+x)(1+y)}\right) ,
\end{aligned}
\end{equation}
where $y_l,y_u=x\cos\gamma \mp \sin\gamma \sqrt{1-x^2} $ and 
$$\eta(x,y)=\sqrt{\sin^2\gamma-(x^2+y^2-2xy\cos\gamma)}$$
is the boundary of the integration area.
The integral can be done analytically (the details of the calculation are provided in Appendix), and the result is 
\begin{widetext}
\begin{equation}\label{ORF_resl}
    \begin{aligned}
        \Gamma^T_{IJ}(f)&=\frac{1}{3}\gamma^T_{IJ}(f)=\frac{1}{16}\bigg\{
        2+\frac{2\cos\gamma}{3}+16\sin^2\frac{\gamma}{2}\mbox{ln}(\sin\frac{\gamma}{2})
        -2i(1-\cos\gamma)\left(\frac{e^{-2i\beta_u}}{\beta_u}-\frac{e^{2i\beta_v}}{\beta_v}\right) \\
        &-\frac{(3+e^{-2i\beta_u})\cos\gamma+1-e^{-2i\beta_u}}{\beta_u^2}-\frac{(3+e^{2i\beta_v})\cos\gamma+1-e^{2i\beta_v}}{\beta_v^2} 
        -2i\cos \gamma\left(\frac{1-e^{-2i\beta_u}}{\beta_u^3}-\frac{1-e^{2i\beta_v}}{\beta_v^3}\right) \\
        &+8\sin^2\frac{\gamma}{2}\left(\mbox{Ei}(-2i\beta_u)+\mbox{Ei}(2i\beta_v)-\mbox{Ei}(-i(\beta_s+\beta_u-\beta_v))-\mbox{Ei}(i(\beta_s-\beta_u+\beta_v))\right) \\
        &+\frac{e^{-i(\beta_u-\beta_v)}}{\beta_u\beta_v}\left[ \cos\beta_s\left(-\frac{1}{2}-3i(\beta_u-\beta_v)-\frac{\beta^2}{\beta_s^2}
        +\frac{3(\beta_u^2-\beta_v^2)^2}{2\beta_s^4}-\frac{i(\beta_u-\beta_v)(\beta_u+\beta_v)^2}{\beta_s^2}\right)\right.\\
        & +\sin\beta_s\left( -\frac{3(\beta_u^2-\beta_v^2)^2}{2\beta_s^5}+\frac{2\beta^2+(\beta_u^2-\beta_v^2)^2}{2\beta_s^3}\right.
        \left. \left. +\frac{1+4\beta_u\beta_v-2i(\beta_u-\beta_v)}{2\beta_s}+\frac{i(\beta_u-\beta_v)(\beta_u+\beta_v)^2}{\beta_s^3}+\frac{7\beta_s}{2}\right) 
        \right]  \bigg\} ,
    \end{aligned}
\end{equation}
\end{widetext}
\begin{figure*}[!t]
    \centering
    \includegraphics[width=0.4\textwidth]{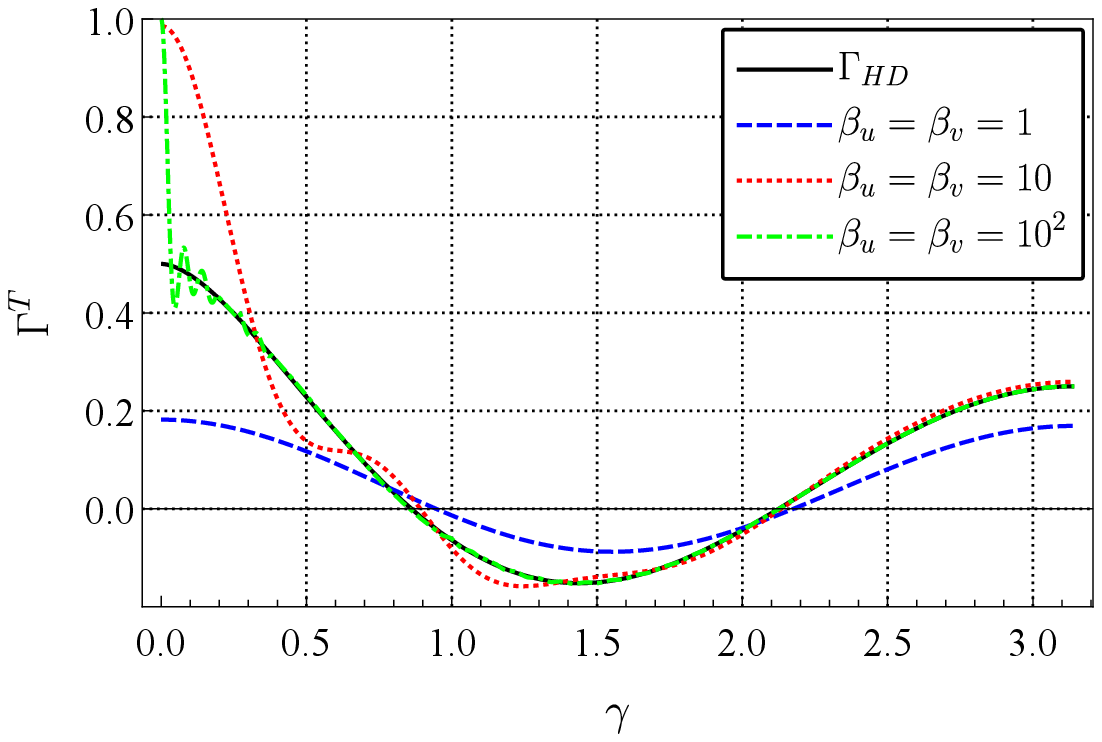}
    \includegraphics[width=0.4\textwidth]{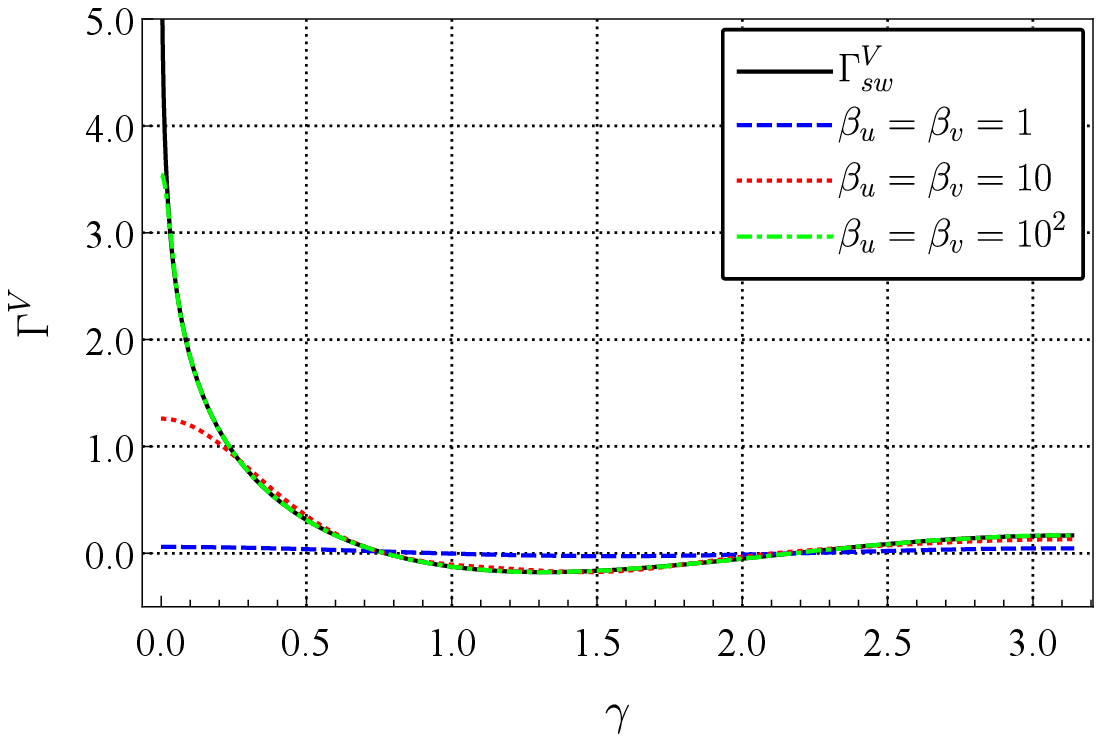} \\
    \includegraphics[width=0.4\textwidth]{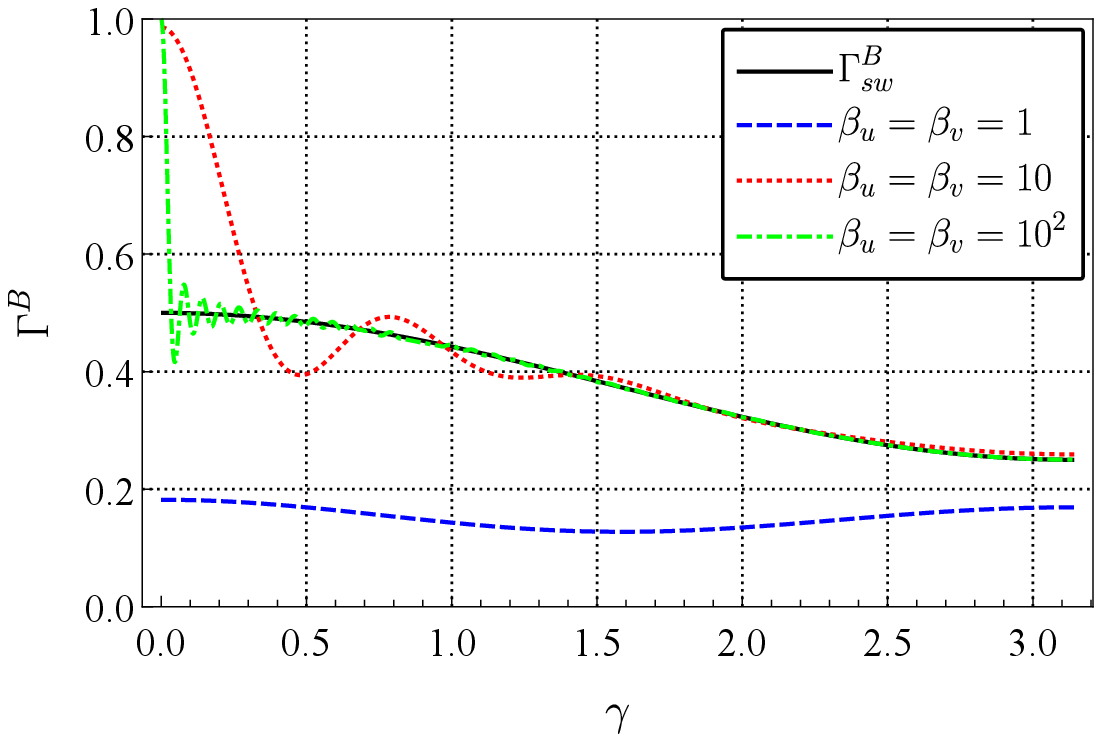}
    \includegraphics[width=0.4\textwidth]{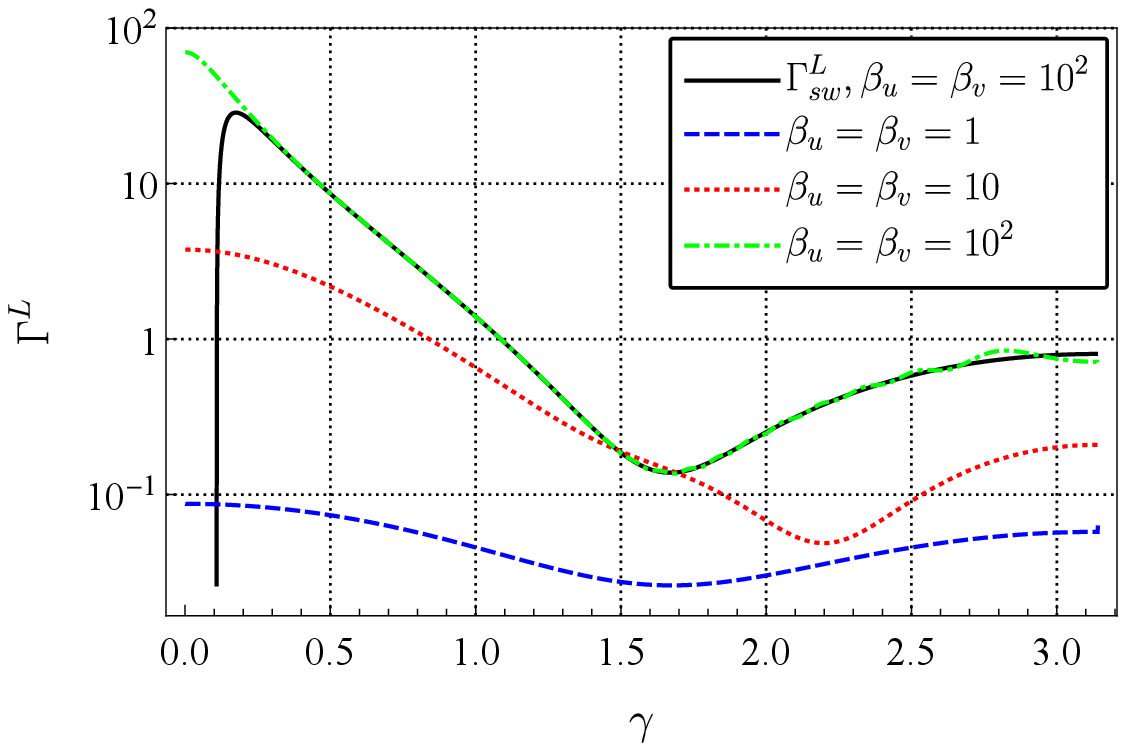}
    \caption{The ORF for different polarizations in the case $L_u=L_v$. The ORF for tensor mode and scalar-breathing mode has been normalized, but not for vector mode and scalar-longitudinal mode. The short-wave approximation expansions (black curves) are also shown for comparison. }\label{fig1}
\end{figure*} 
\begin{figure*}[!t]
    \centering
    \includegraphics[width=0.4\textwidth]{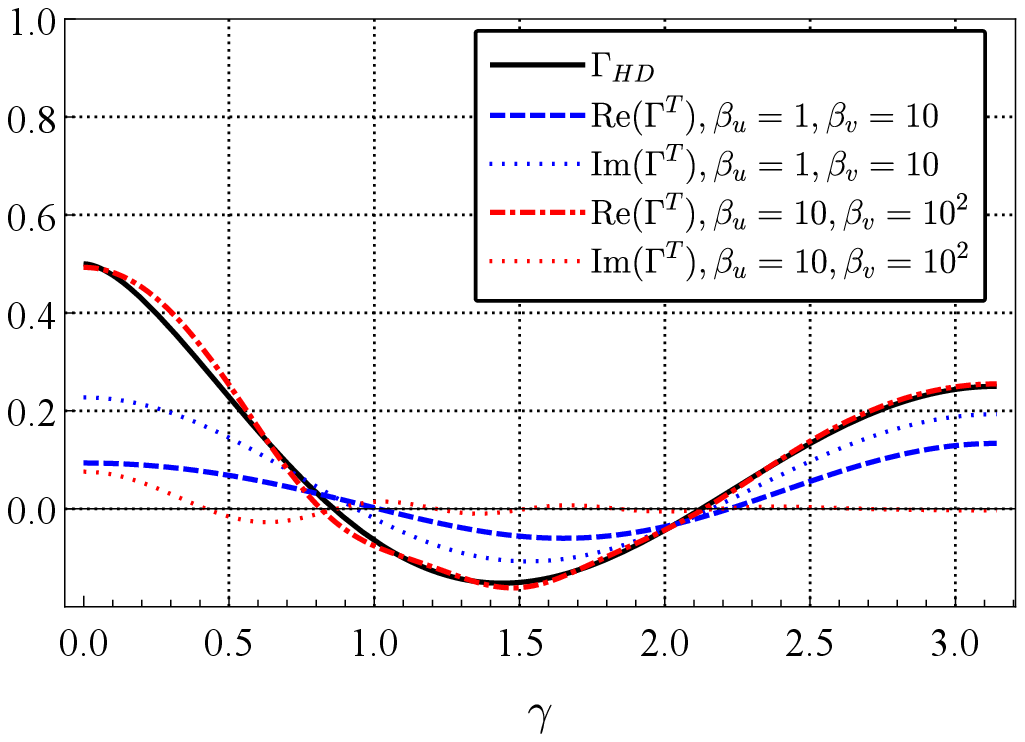}
    \includegraphics[width=0.4\textwidth]{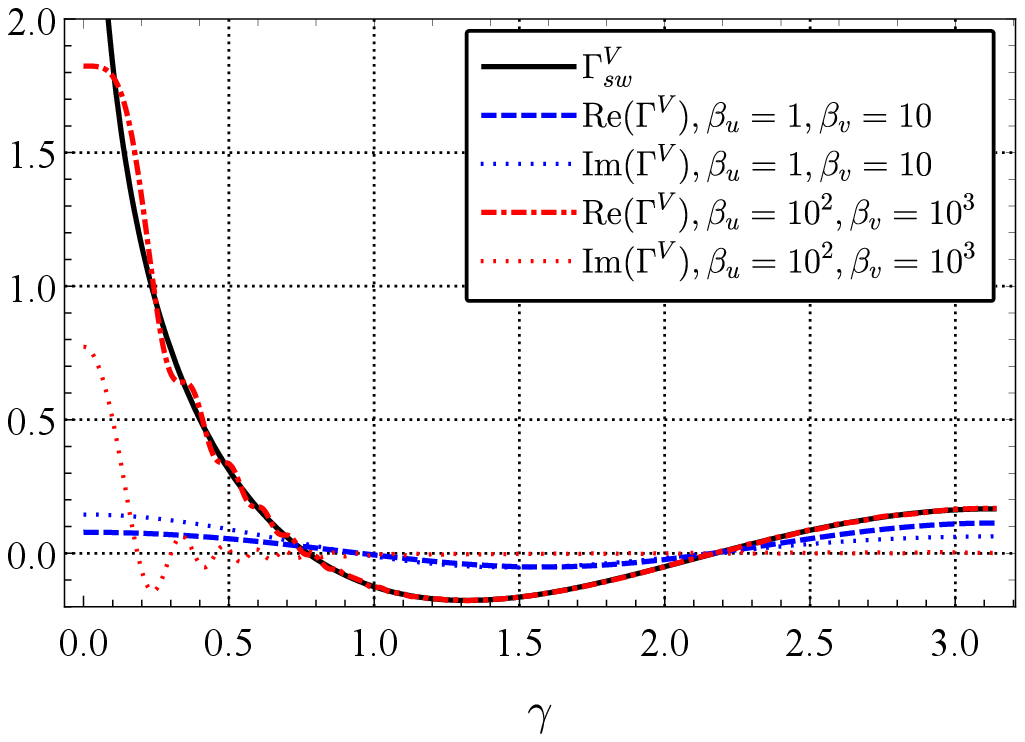} \\
    \includegraphics[width=0.4\textwidth]{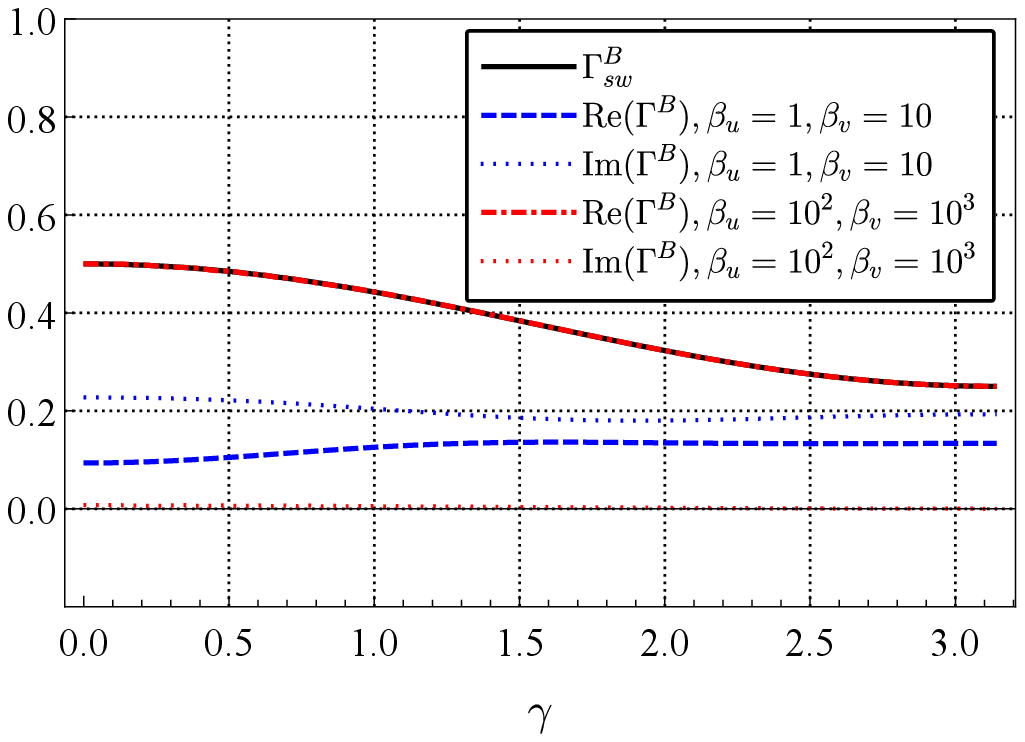}
    \includegraphics[width=0.4\textwidth]{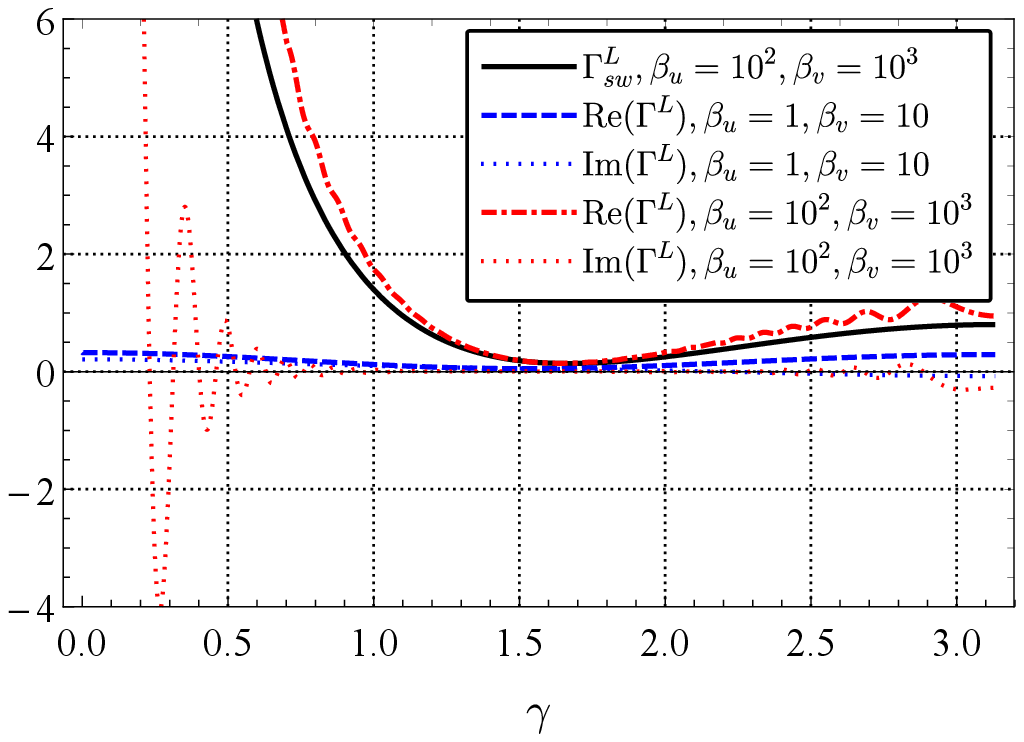}
    \caption{The ORF for different polarizations in the case $L_u \neq L_v$. }\label{fig2}
\end{figure*}
where 
$$\beta=\sqrt{\beta_u^2+\beta_v^2} ,$$
$$\beta_s=\sqrt{\beta_u^2+\beta_v^2-2\beta_u\beta_v\cos \gamma} ,$$
and $\text{Ei}(z)=-\int_{-z}^{\infty}e^{-t}/tdt$ is the exponential integral function.
$\gamma^T_{IJ}(f)$ is the normalized ORF, so that $\gamma^T_{IJ}(f)=1$ for a single pulsar.
The normalization holds only for the short-wave approximation.
Using the short-wave approximation $\beta_u,\beta_v \gg 1$, Eq. (\ref{ORF_resl}) will reduce to the HD function \cite{1983apj_HD_curve},
\begin{equation}\label{HD}
    \gamma_{HD}(\gamma)=\frac{3}{2}\left(\frac{1-\cos \gamma}{2}\right)\ln \frac{1-\cos \gamma}{2}-\frac{1-\cos \gamma}{8}+\frac{1}{2} .
\end{equation}
In addition, it can also accurately converge to 1 when the two pulsars tend to coincide, because the exponential terms is included.

Since no approximations are used, the analytic expressions agree exactly with the values obtained by direct numerical integration.
When the distances of two pulsars are the same ($\beta_u=\beta_v $), the ORF is a real function.
As shown in Fig. \ref{fig1}, our result is consistent with the HD curve as $\beta_u$ increase, and the curve approaches 1 smoothly when the two pulsars are very close.
For the long-wave wavelengths, $\lim_{\gamma \rightarrow 0}\gamma^T_{IJ}$ is a value less than 0.5, which depends on $\beta_u$.
However, the two pulsars are usually at different distances in reality, which implies that the ORF is a complex function and satisfies that $\Gamma_{IJ}(-f)=\Gamma^*_{IJ}(f)$. 
As shown in Fig. \ref{fig2}, for different distances of pulsars ($\beta_u \neq \beta_v $), the imaginary part of the ORF depends on the ratio of $\beta_u$ and $\beta_v $.
The greater the difference between them, the greater the value of the imaginary part.
But when both $\beta_u$ and $\beta_v $ are very large, the imaginary part is close to 0, no matter how large their ratio is.
And the real part is tend to HD function in this case.

We show the behavior in Fig. \ref{fig3} when the angle between the two pulsars is small.
If $\beta_u=\beta_v $, $\gamma^T_{IJ}$ decays from 1 to 0.5 faster and faster as $\beta_u$ increases.
From their similar curves, we can conclude that the ORF for two nearby pulsars depends on their distance from each other.
For pulsars in the same direction $\gamma=0$,
\begin{equation}
    \begin{aligned}\label{ORF_approx}
    \lim_{\gamma\rightarrow 0}&\gamma^T_{IJ}(f)=\frac{1}{2}-\frac{3}{4\beta_u^2}-\frac{3}{4\beta_v^2}+\frac{3}{4(\beta_u-\beta_v)^2} \\
    &-i\frac{3}{8}\left(\frac{1-e^{-2i\beta_u}}{\beta_u^3}-\frac{1-e^{2i\beta_v}}{\beta_v^3}\right)\\
    &-e^{-i(\beta_u-\beta_v)}\frac{3\sin(\beta_u-\beta_v)}{4(\beta_u-\beta_v)^3}-\frac{3i(\beta_u^2-\beta_u\beta_v+\beta_v^2)}{4\beta_u\beta_v(\beta_u-\beta_v)} .
    \end{aligned}
\end{equation}
In addition, we noticed that $\lim_{\gamma \rightarrow 0}\gamma^T_{IJ} \neq 1$ when $\beta_u$ is not very large.
Actually, the auto-correlated response for the tensor mode of a single detector is 
\begin{equation}\label{gamma_T_II}
    \gamma^T_{II}(f)=\lim_{\substack{\gamma\rightarrow 0 \\ \beta_v\rightarrow \beta_u}}\gamma^T_{IJ}(f)=1-\frac{3}{2\beta_u^2}+\frac{3\sin2\beta_u}{4\beta_u^3} .
\end{equation}
It dose converge to 1 for $\beta_u \gg 1$, but gradually decreases to 0 as $\beta_u$ decreases.
The normalization expression $\gamma^T_{IJ}(f)$ is only valid for short-wave approximation.

\subsection{Vector mode}
The ORF of the vector mode isotropic is
\begin{widetext}
\begin{equation}\label{ORF_V}
    \begin{aligned}
    \Gamma_{IJ}^V(f)=\frac{1}{32\pi}\int d^2\Omega_{\hat{n}}
    \frac{(1-e^{-i\beta_u (1+\sin\theta\cos\underset{\sim}{\phi})})}
    {(1+\sin\theta\cos\underset{\sim}{\phi}) }
    \frac{(1-e^{i\beta_v (1+\sin\theta\cos\tilde{\phi})})} 
    {(1+\sin\theta\cos\tilde{\phi}) }
    4\left( \sin^2\theta\cos\underset{\sim}{\phi}\cos\tilde{\phi}
    (\cos\gamma-\sin^2\theta\cos\underset{\sim}{\phi}\cos\tilde{\phi})    \right) .
    \end{aligned}
\end{equation}
Changing the variables, then we get 
\begin{equation}\label{ORF_V_xy}
    \begin{aligned}
    \Gamma&^V_{IJ}(f)=\frac{1}{16\pi}\int_{-1}^1dx\int_{y_l}^{y_u}dy
    (1-e^{-i\beta_u (1+x)})
    (1-e^{i\beta_v (1+y)}) 
    \frac{4xy(\cos\gamma-xy)}{\eta(x,y)(1+x)(1+y)},
\end{aligned}
\end{equation}
Finishing the integral, and we find that
\begin{equation}\label{ORF_V_resl}
    \begin{aligned}
        \Gamma_{IJ}^V(f)&=-\frac{1}{2} -\frac{2 \cos \gamma }{3}
        -\mbox{ln} \sin \frac{\gamma}{2}
         +\frac{i \left(2 \cos \gamma +1-e^{2 i \beta_{v}}\right)}{4 \beta_{v}}
        -\frac{i \left(2 \cos \gamma +1-e^{-2 i \beta_{u}}\right)}{4\beta_{u}} \\
        &+\frac{\left(3+e^{-2 i \beta_{u}}\right) \cos \gamma -e^{-2 i
           \beta_{u}}+1}{4 \beta_{u}^2}
        +\frac{\left(3+e^{2 i {\beta_v}}\right) \cos \gamma -e^{2 i \beta_{v}}+1}{4 \beta_{v}^2}
        +\frac{i\cos \gamma}{2}\left(\frac{1-e^{-2 i \beta_{u}}}{\beta_{u}^3} -\frac{1-e^{2 i \beta_{v}}}{\beta_{v}^3}\right)\\
        &+\frac{1}{2} e^{-i (\beta_{u}-\beta_{v})} 
        \left\{ 
        \cos \beta_{s}
           \left(\frac{3 \beta_{u} \beta_{v} \sin ^2\gamma }{\beta_{s}^4}
        +\frac{-2 \cos \gamma +i (\beta_{u}-\beta_{v}) (\cos \gamma+1)}{\beta_{s}^2}
        -i \left(\frac{1}{\beta_{u}}-\frac{1}{\beta_{v}}\right)\right) 
        \right.\\
        &\left.
        +\sin \beta_{s} 
        \left(-\frac{3 \beta_{u}\beta_{v} \sin ^2\gamma }{\beta_{s}^5}
        +\frac{\beta_{u}\beta_{v} \sin ^2\gamma -i (\beta_{u}-\beta_{v}) (\cos \gamma+1)
        +2 \cos \gamma }{\beta_{s}^3}
        -\frac{2 \cos \gamma +1}{\beta_{s}}
        -\frac{\beta_{s}}{\beta_{u}\beta_{v}} \right)
         \right\}\\
        &+\frac{1}{2} \left(\mbox{Ei}(-i (\beta_{s}+\beta_{u}-{\beta_{
           v}}))+\mbox{Ei}(i (\beta_{s}-\beta_{u}+\beta_{v}))-\mbox{Ei}(-2 i
           \beta_{u})-\mbox{Ei}(2 i \beta_{v})\right) .
        \end{aligned}
\end{equation}
\end{widetext}
The ORF for vector mode cannot be normalized, because $\Gamma_{IJ}^V(\gamma=0)$ increase with $\beta_u$.
The expression with using short-wave approximation is \cite{2008apj_Lee_ORF_nonGR}
\begin{equation}\label{ORF_V_sw}
    \Gamma^V_{sw}(\gamma)=-\frac{1}{2}-\frac{2}{3}\cos \gamma-\ln (\sin\frac{\gamma}{2}),
\end{equation}
where the angular separation $\gamma$ is assumed to be not small.
In the limit $\gamma \rightarrow 0$, Eq. (\ref{ORF_V_resl}) will give the exact value while the short-wave expression $\Gamma^V_{sw}$ diverges.
The auto-correlated response for vector mode of a single detector is 
\begin{equation}\label{gamma_V_II}
    \begin{aligned}
    \Gamma^V_{II}(f)=&\lim_{\substack{\gamma\rightarrow 0 \\ \beta_v\rightarrow \beta_u}}\Gamma^V_{IJ}(f)
    =-\frac{7}{3}+\gamma_E+\text{ln}(2\beta_u)-\text{Ci}(2\beta_u) \\
    &+\frac{2}{\beta^2_u}+\frac{(\beta^2_u-2)\cos\beta_u\sin\beta_u}{\beta^3_u} ,
    \end{aligned}
\end{equation}
where $\gamma_E$ is the Euler constant and 
$$\text{Ci}(z)=-\int_z^{\infty}\cos t/t dt $$ is the cosine-integral function.
In the limit $\gamma \rightarrow 0$, $\Gamma^V$ grows logarithmically with $2\beta_u$.
It is consistent with Eq.(A42) in \cite{2008apj_Lee_ORF_nonGR}.

\begin{figure*}[!t]
    \flushleft
        \hspace{15pt}\includegraphics[width=0.38\textwidth]{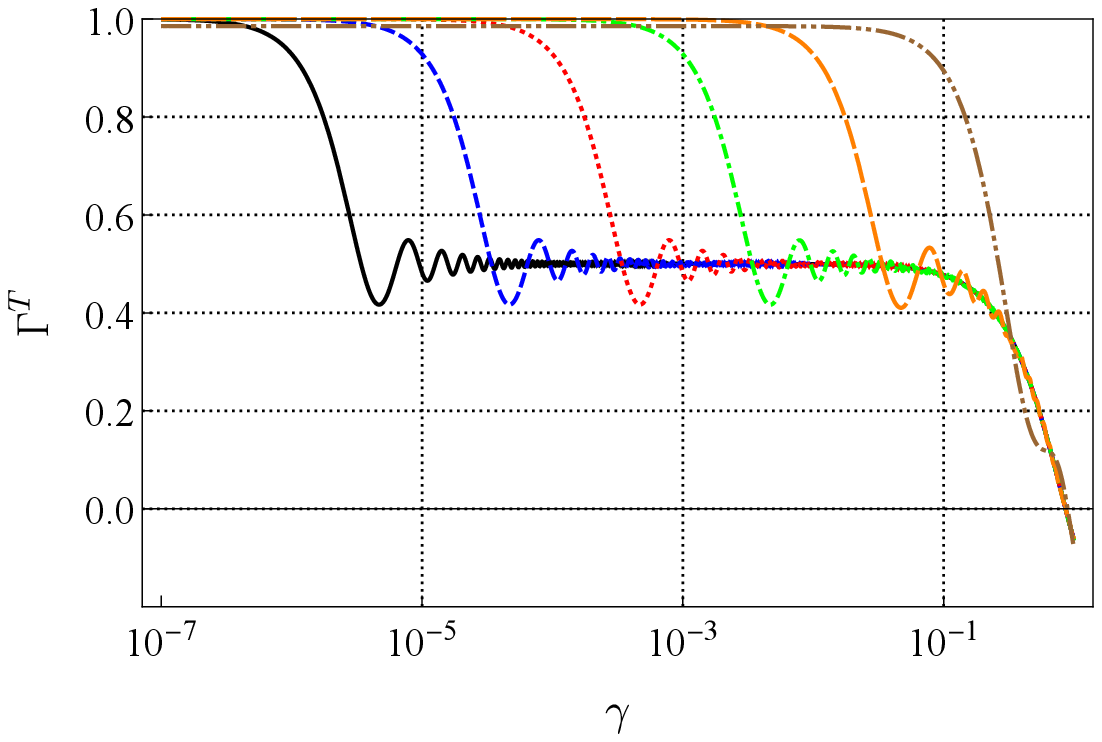}
        \includegraphics[width=0.38\textwidth]{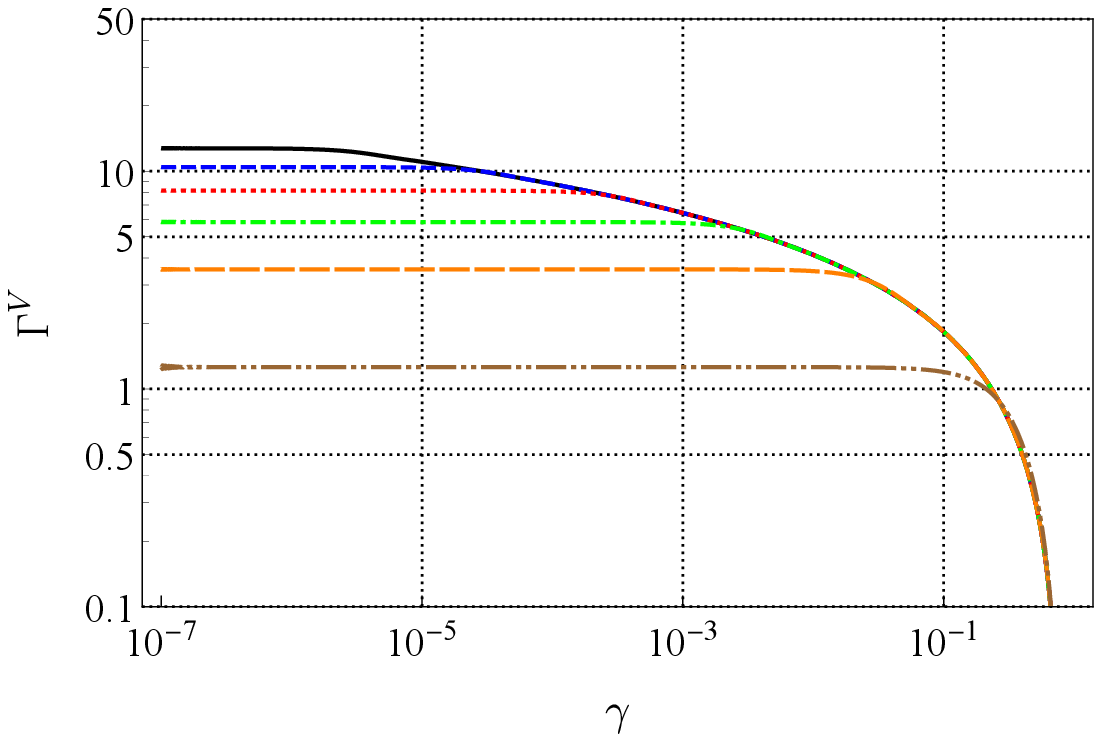} 
        \hspace{10pt}\includegraphics[width=0.13\textwidth]{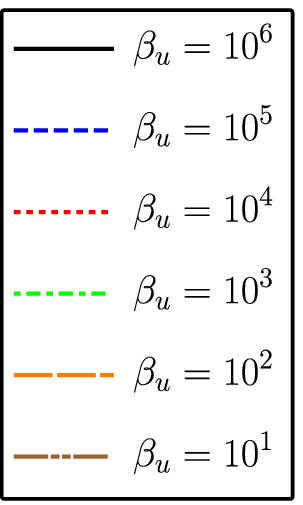}\\
        \hspace{15pt}\includegraphics[width=0.38\textwidth]{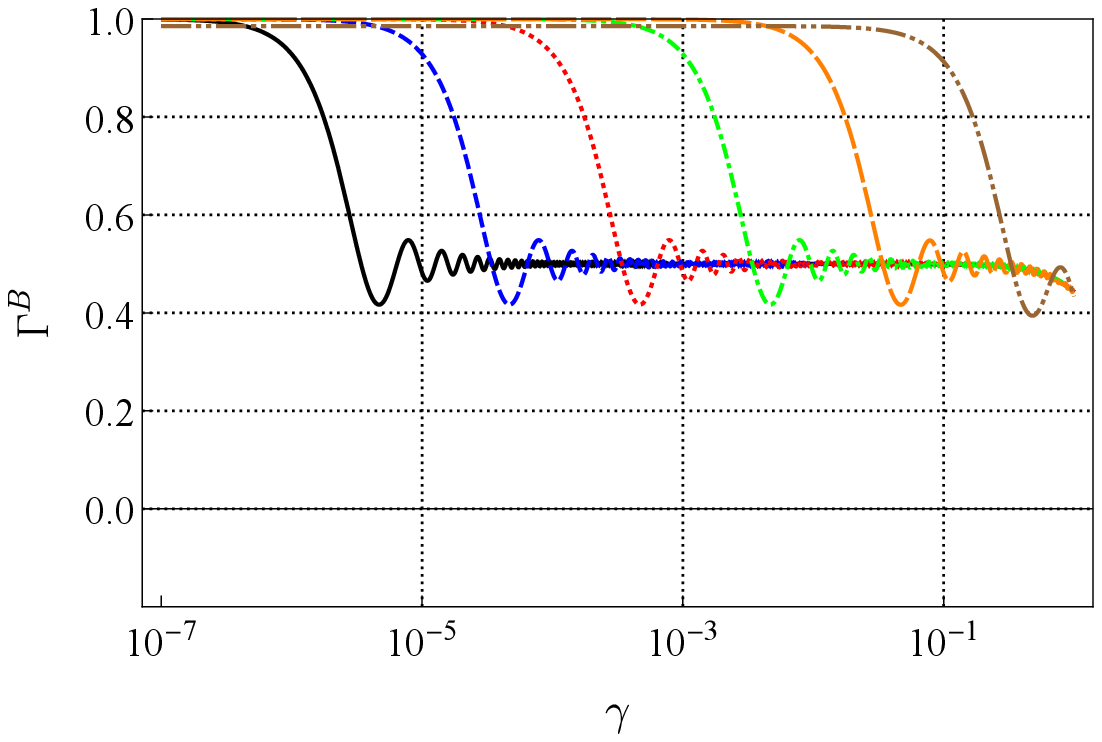}
        \includegraphics[width=0.38\textwidth]{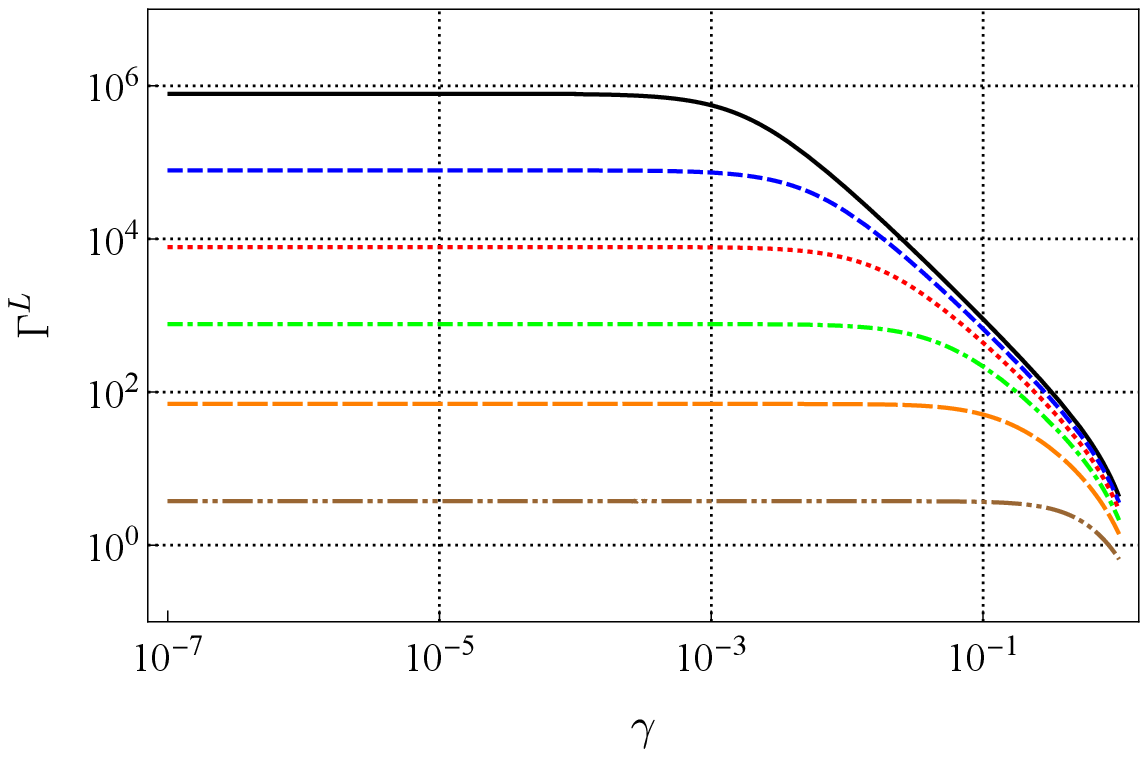}
    \caption{The behavior of ORF for small angles when $\beta_v=\beta_u$.}\label{fig3}
\end{figure*}

The behaviors of ORF for vector mode are shown in Fig. \ref{fig1} and Fig. \ref{fig2} in the case $\beta_u=\beta_v$ and $\beta_u \neq \beta_v$ respectively.
For two pulsars at the same distance, the ORF for vector mode gradually approaches the short-wave expression as $\beta_u$ increases.
From the short-wave expression (\ref{ORF_V_sw}), we know that $\Gamma^V_{sw}(\gamma)$ diverges at $\gamma=0$.
However, at $\gamma=0$, the full analytic expression tends to have finite values $\Gamma^V_{II}(\beta_u)$, which grows logarithmically with $2\beta_u$.
For two pulsars at different distances, the ORF for vector mode is complex.
The real part will also gradually approaches the short-wave expression for not too small of $\gamma$.
However, when $\gamma$ is very small, the imaginary part is not negligible compared to the real part even 
$\beta_u$ and $\beta_v$ are pretty large.
The small angle behavior of vector mode is quite different from the tensor mode.
As shown in Fig. \ref{fig3}, the ORF for vector mode converges to a finite value, which depends on $\beta_u$ and increases with $\beta_u$ logarithmically.
There is also no fast halving decay, and the transition of the curve is smooth.

\subsection{Scalar-breathing mode}
The ORF of scalar-breathing mode is 
\begin{widetext}
\begin{equation}\label{ORF_Sb}
    \begin{aligned}
    \Gamma_{IJ}^B(f)=&\frac{1}{16\pi}\int d^2\Omega_{\hat{n}}
    \frac{(1-e^{-i\beta_u (1+\sin\theta\cos\underset{\sim}{\phi})})}
    {(1+\sin\theta\cos\underset{\sim}{\phi}) }
    \frac{(1-e^{i\beta_v (1+\sin\theta\cos\tilde{\phi})})} 
    {(1+\sin\theta\cos\tilde{\phi}) } 
    (1-\sin^2\theta\cos^2\underset{\sim}{\phi})(1-\sin^2\theta\cos^2\tilde{\phi}) \\
    =&\frac{1}{8\pi}\int_{-1}^1dx\int_{y_l}^{y_u}dy
    (1-e^{-i\beta_u (1+x)})
    (1-e^{i\beta_v (1+y)}) 
     \frac{(1-x)(1-y)}{\eta(x,y)} .
    \end{aligned}
\end{equation}
The normalized analytic expression of ORF for scalar-breathing mode is 
\begin{equation}\label{ORF_B_resl}
    \begin{aligned}
    \gamma_{IJ}^B(f)&=\frac{3}{2}\Gamma_{IJ}^B(f)=\frac{3}{16} \bigg \{ 2+\frac{2 \cos \gamma }{3} 
    +2 i(1+\cos\gamma) \left(\frac{1}{\beta_u}-\frac{1}{\beta_v}\right)
    -2i \cos \gamma\left(\frac{1-e^{-2 i \beta_u}}{\beta_u^3}-\frac{1-e^{2 i\beta_v}}{\beta_v^3}\right) \\
    &+\frac{e^{-2 i \beta_u}-1 -\left(3+e^{-2 i\beta_u}\right) \cos \gamma }{\beta_u^2}
    +\frac{e^{2 i\beta_v}-1 -\left(3+e^{2 i \beta_v}\right) \cos \gamma }{\beta_v^2}\\
    &+\frac{e^{-i (\beta_u-\beta_v)}}{\beta_u \beta_v}
    \left[\sin\beta_s \left(\frac{2 \beta ^2+\left(\beta_u^2-\beta_v^2\right)^2}{2 \beta_s^3}
    -\frac{3 \left(\beta_u^2-\beta_v^2\right)^2}{2 \beta_s^5} \right. \right. \\
    &\left.+\frac{i (\beta_u-\beta_v)(\beta_u+\beta_v)^2}{\beta_s^3}
    +\frac{-2 i ({\beta_u}-\beta_v)+4 \beta_u \beta_v+1}{2 \beta_s}
    -\frac{\beta_s}{2}\right) \\
    &\left.+\cos \beta_s \left(-\frac{\beta^2}{\beta_s^2}+\frac{3 \left(\beta_u^2-\beta_v^2\right)^2}{2 \beta_s^4}
    -\frac{i (\beta_u-\beta_v)(\beta_u+\beta_v)^2}{\beta_s^2}+i (\beta_u-\beta_v)-\frac{1}{2}\right)\right] \bigg \} .
    \end{aligned}
\end{equation}
\end{widetext}
The short-wave expression is \cite{2008apj_Lee_ORF_nonGR}
\begin{equation}
    \gamma^{B}_{sw}(\gamma)=\frac{3}{8}+\frac{1}{8}\cos\gamma .
\end{equation}
The auto-correlated response for the scalar-breathing mode of a single detector is the same as that of the tensor mode,
\begin{equation}\label{gamma_B_II}
    \gamma^B_{II}(f)=\lim_{\substack{\gamma\rightarrow 0 \\ \beta_v\rightarrow \beta_u}}\gamma^B_{IJ}(f)=1-\frac{3}{2\beta_u^2}+\frac{3\sin2\beta_u}{4\beta_u^3} .
\end{equation}
But the actual responses satisfy that $\Gamma^B_{II}(f)=1/2 \Gamma^T_{II}(f)$, because their normalization factors are different.
Overall, the ORF for the scalar-breathing mode is very similar to that for the tensor mode, probably because they are both transverse modes.
In particular, it should be noticed that $\Gamma_{IJ}^B(f) \ge 0$ for any configuration of pulsar pairs.

\subsection{Scalar-longitudinal mode}
The ORF of scalar-longitudinal mode is 
\begin{widetext}
\begin{equation}\label{ORF_Sl}
    \begin{aligned}
    \Gamma_{IJ}^L(f)=&\frac{1}{16\pi}\int d^2\Omega_{\hat{n}}
    \frac{(1-e^{-i\beta_u (1+\sin\theta\cos\underset{\sim}{\phi})})}
    {(1+\sin\theta\cos\underset{\sim}{\phi}) }
    \frac{(1-e^{i\beta_v (1+\sin\theta\cos\tilde{\phi})})} 
    {(1+\sin\theta\cos\tilde{\phi}) } 
     2(\sin^4\theta\cos^2\underset{\sim}{\phi}\cos^2\tilde{\phi}) \\
    =&\frac{1}{8\pi}\int_{-1}^1dx\int_{y_l}^{y_u}dy
    (1-e^{-i\beta_u (1+x)})
     (1-e^{i\beta_v (1+y)}) 
     \frac{2 x^2 y^2}{\eta(x,y)(1+x)(1+y)} .
    \end{aligned}
\end{equation}
The analytic expression of ORF for scalar-longitudinal mode is 
\begin{equation}\label{ORF_L_resl}
    \begin{aligned}
        \Gamma_{IJ}^L(f)&=\frac{1}{8}\biggl\{4 + \frac{28 \cos \gamma }{3} + \csc ^2\frac{\gamma }{2}
        \left(4\ln (\sin\frac{\gamma }{2}) + 2\gamma_E \cos ^2\gamma + \cos ^2\gamma \ln (4 \beta_u\beta_v) \right) \\
        &+2 i \left(\frac{1}{\beta_u}-\frac{1}{\beta_v}\right) \left(\frac{1}{2}
        \cos \beta_s e^{-i (\beta_u-\beta_v)} \left(4-\csc^2\frac{\gamma }{2}-\sec ^2\frac{\gamma }{2}\right)
        +3 \cos\gamma +1\right)\\
        &-2 i \left(\frac{e^{-2 i \beta_u}}{\beta_u}-\frac{e^{2 i \beta_v}}{\beta_v}\right) (\cos \gamma +1)
        -4 i \cos \gamma \left(\frac{1-e^{-2 i \beta_u}}{\beta_u^3}
        -\frac{1-e^{2 i\beta_v}}{\beta_v^3}\right)\\
        &+\frac{-2\left(3+ e^{-2 i \beta_u}\right) \cos \gamma +2 e^{-2 i \beta_u}-2}{\beta_u^2}
        +\frac{-2 \left(3+e^{2 i \beta_v}\right) \cos\gamma +2 e^{2 i \beta_v}-2}{\beta_v^2}\\
        &-\frac{1}{2} (\cos 2 \gamma -3) \csc ^2\frac{\gamma }{2}\left(\mbox{Ei}(-2 i \beta_u)+\mbox{Ei}(2 i \beta_v)\right)
        -2 \csc ^2\frac{\gamma }{2} 
        \left(\mbox{Ei}(-i (\beta_s+\beta_u-\beta_v))+\mbox{Ei}(i (\beta_s-\beta_u+\beta_v)) \right)\\
        &-\cos ^2\gamma  \csc ^2\frac{\gamma }{2}
        \left[e^{-2 i \beta_u\sin ^2\frac{\gamma }{2}} \left(\mbox{Ei}(-i \beta_u (\cos \gamma -1))+\mbox{Ei}(-i \beta_u (\cos \gamma +1))\right. \right.\\
        &\left.-\mbox{Ei}(i (\beta_s+\beta_v-\beta_u \cos \gamma ))-\mbox{Ei}(-i (\beta_s-\beta_v+\beta_u \cos \gamma ))
         \right)\\
        &\left. +e^{2 i \beta_v \sin^2\frac{\gamma }{2}}
        \left(\mbox{Ei}(i \beta_v (\cos \gamma -1))+\mbox{Ei}(i \beta_v (\cos \gamma +1)) \right.
        \left.-\mbox{Ei}(-i (\beta_s+\beta_u-\beta_v \cos \gamma ))-\mbox{Ei}(i (\beta_s-\beta_u+\beta_v \cos \gamma ))\right)\right]\\
        &+e^{-i (\beta_u-\beta_v)}\left[\cos \beta_s
        \left(-\frac{12\beta_u \beta_v \sin ^2\gamma }{\beta_s^4}
        -\frac{-8 \cos\gamma +4 i (\beta_u-\beta_v) (\cos \gamma +1)}{\beta_s^2}
        -\frac{4 i (\beta_u-\beta_v) \csc ^2\gamma }
       {\beta_u \beta_v}\right) \right.\\
        &\left. +\sin \beta_s \left(\frac{12 \beta_u \beta_v \sin ^2\gamma}{\beta_s^5}
        -\frac{4 \beta_u \beta_v \sin ^2\gamma -4 i(\beta_u-\beta_v) (\cos \gamma +1)+8 \cos \gamma }{\beta_s^3}
        +\frac{4 \beta_s}{\beta_u \beta_v}+\frac{12 \cos\gamma +4}{\beta_s}\right) \right] \biggl\} .
    \end{aligned}
\end{equation}
\end{widetext}
Like the vector mode, the ORF of the scalar-longitudinal mode cannot be normalized.
The auto-correlated response for the scalar-longitudinal mode of a single detector is
\begin{equation}\label{ORF_L_II}
    \begin{aligned}
        \Gamma^L_{II}(f)=&\lim_{\substack{\gamma\rightarrow 0 \\ \beta_v\rightarrow \beta_u}}\Gamma^L_{IJ}(f)
        =\frac{37}{12}-2\gamma_E-2\ln (2\beta_u) \\
        &+\frac{1}{2}\beta_u \mbox{Si} (2\beta_u) +\frac{1}{4}\cos(2\beta_u) \\
        &+2\mbox{Ci}(2\beta_u)-\frac{\sin(2\beta_u)}{\beta_u}-\frac{2}{\beta^2_u}+\frac{\sin(2\beta_u)}{\beta^3_u} .
    \end{aligned}
\end{equation}
For short wavelengths, $ \Gamma^L_{II} \propto \frac{\pi}{4}\beta_u $ and it grows faster than the logarithmically growing vector mode as $\beta_u$ increases.
This is consistent with Eq.(A36) in \cite{2008apj_Lee_ORF_nonGR} and Eq.(41) in \cite{2012prd_ORF_nonGR}.

For the scalar-longitudinal mode, there is no analytic expression of ORF even using the short-wave approximation.
Dropping the exponential terms and directly calculating integral leads to divergence \cite{2017Review_detection_GWB}.
The exponential terms must be included to overcome the divergence.
However, we can obtain a short-wave expression from Eq. (\ref{ORF_L_resl}),
\begin{equation}\label{ORF_L_sw}
    \begin{aligned}
    \Gamma&^L_{sw}(f)=\frac{1}{2}+\frac{7}{6}\cos \gamma \\
     &+\frac{1}{8}\csc^2 \frac{\gamma}{2}
    \left(\cos^2\gamma \left(2\gamma_E+\ln (4\beta_u\beta_v)\right) +4\ln \sin\frac{\gamma}{2}\right) .
    \end{aligned}
\end{equation}
As shown in Fig. \ref{fig1}, the approximate expression works well except that $\gamma$ is close to $0$ and $\pi$.
When the distance difference between the two pulsars is large, the accuracy of the approximate expression becomes worse.
The ORF for scalar-longitudinal mode of pulsars at different distances is shown in Fig. \ref{fig2}.
When $\gamma$ is small, the imaginary part is comparable to the real part for large $\beta_u$ and $\beta_v$.
And the short-wave expression (\ref{ORF_L_sw}) underperforms in this case.
In addition, we notice that $\Gamma^L_{IJ}$ also seems to increase with $\beta_u$ at $\gamma=\pi$.
In fact, it can be shown that this is indeed the case and it grows logarithmically.
$\Gamma^L_{IJ}$ increases linearity with $\beta_u$ at $\gamma=0$ and logarithmically at $\gamma=\pi$.
So what is the behavior between 0 and $\pi$?
$\Gamma^L_{IJ}$ increases with $\beta_u$ except at $\gamma=\pi/2$, where it tends to be a constant.
Explicitly,
\begin{equation}
    \lim_{\beta_u,\beta_v\rightarrow \infty }\Gamma^L_{IJ}(\gamma=\pi/2)=\frac{1}{2}(1-\ln2) .
\end{equation}
The small angle behavior is shown in Fig. \ref{fig3}. 
As $\gamma$ decreases, it tends to a constant, which is consistent with Eq. (\ref{ORF_L_II}).
There is also no fast halving decay, which implies that the exponential terms accounts a non-negligible proportion of the integral.

\section{Discuss and conclusions\label{sec4}}
\begin{figure*}[!t]
    \centering
    \includegraphics[width=0.42\textwidth]{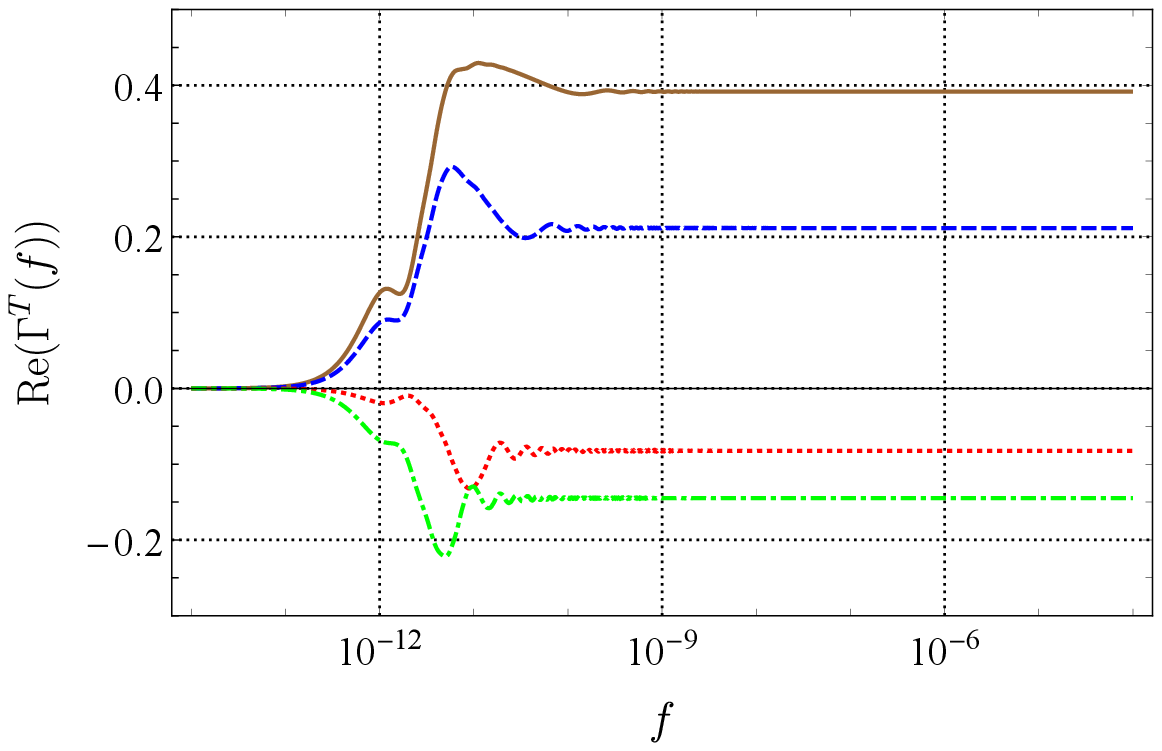}
    \includegraphics[width=0.42\textwidth]{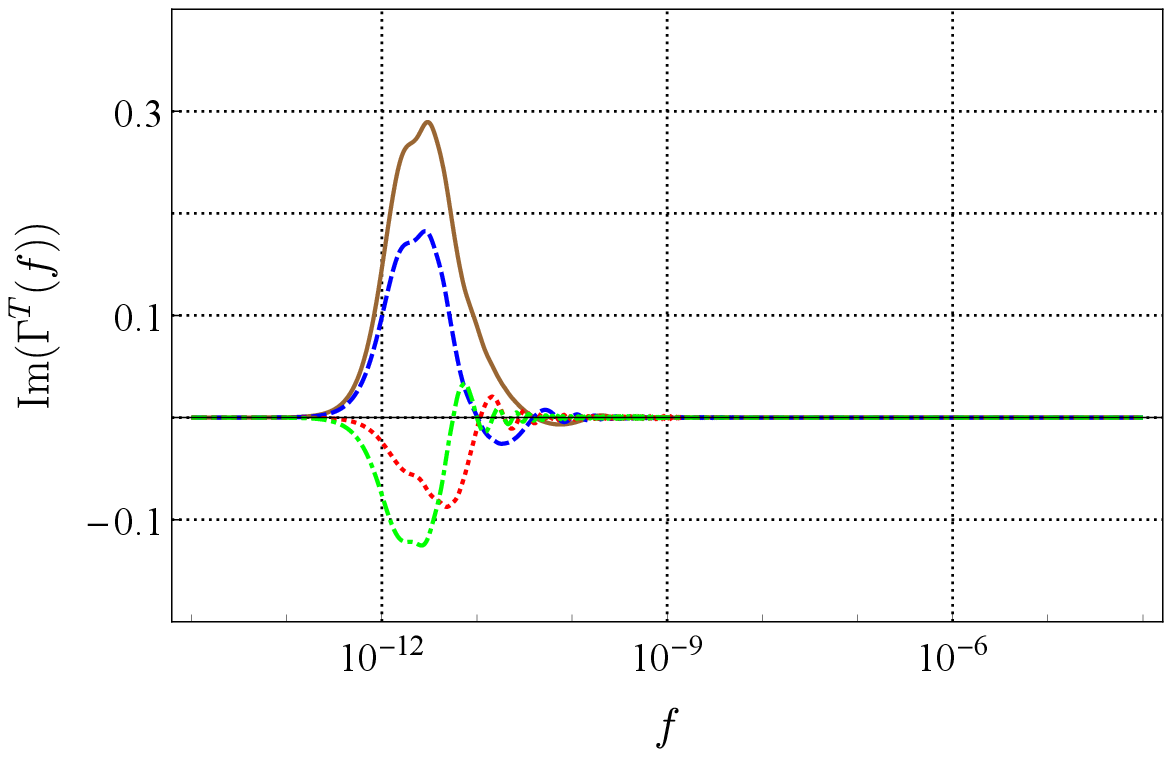} \\
    \includegraphics[width=0.42\textwidth]{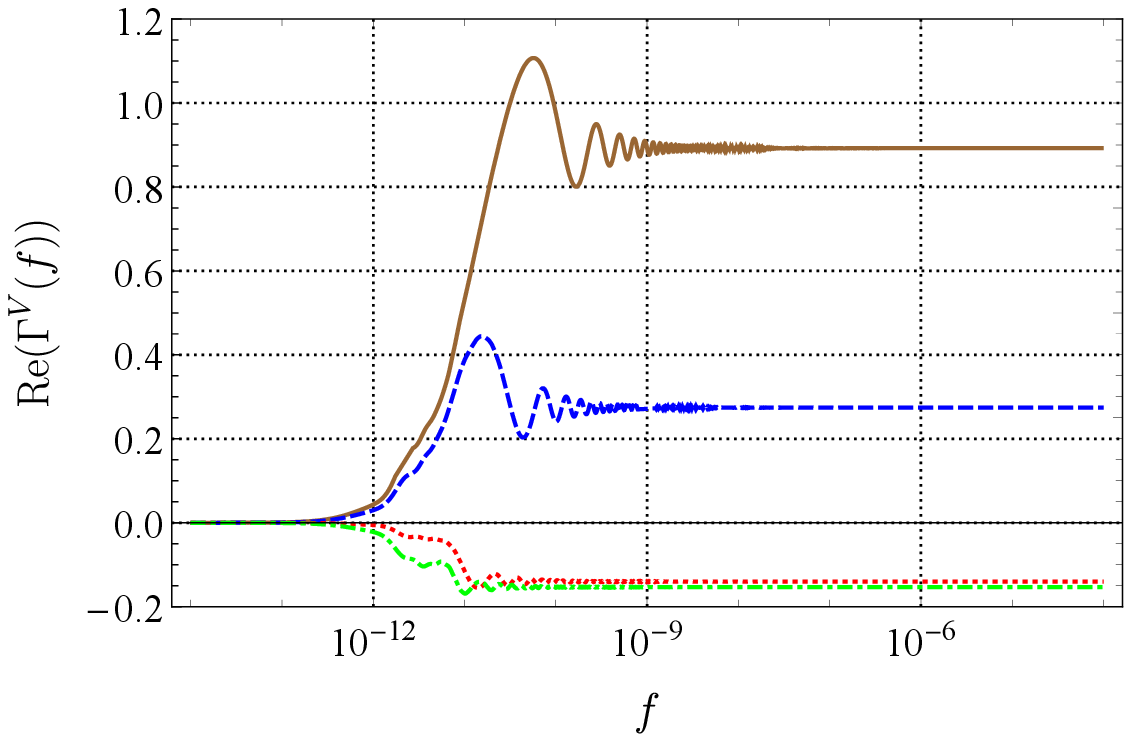}
    \includegraphics[width=0.42\textwidth]{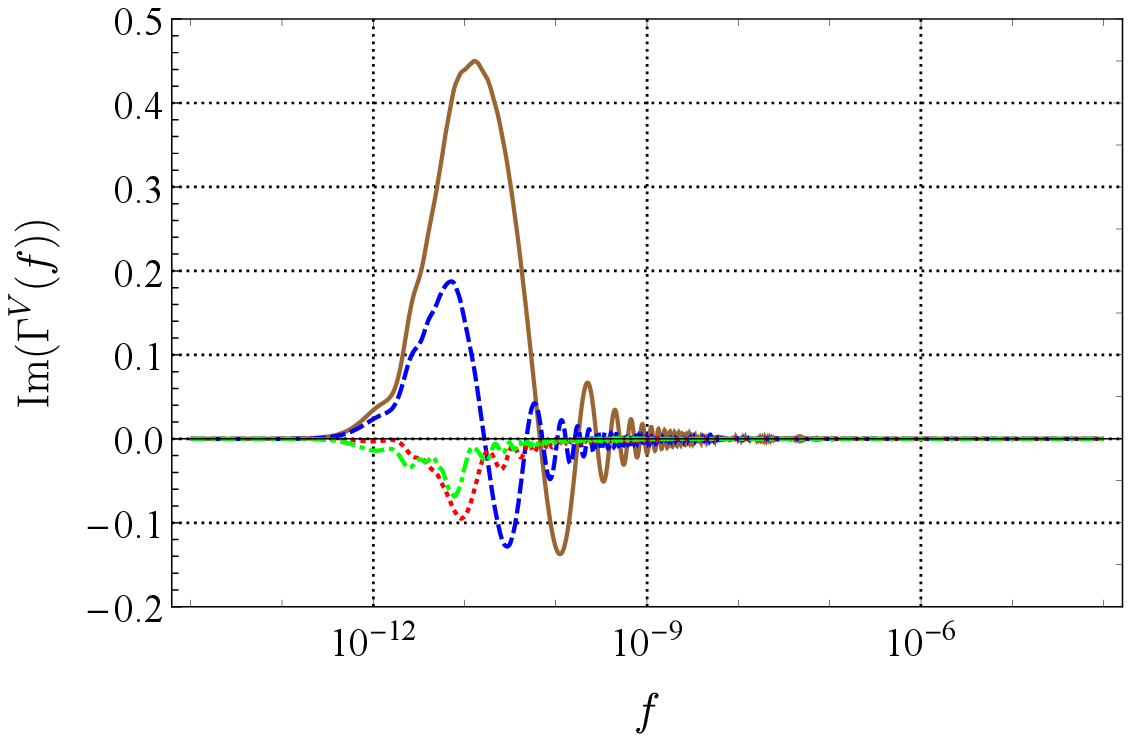} \\
    \includegraphics[width=0.42\textwidth]{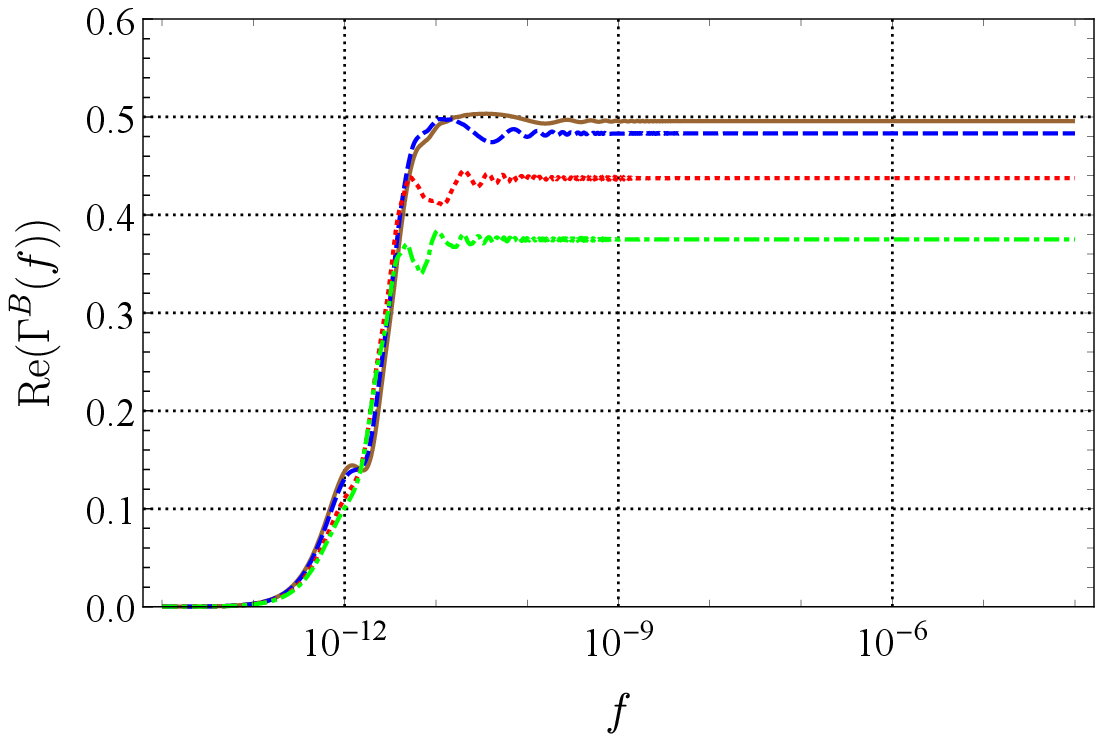}
    \includegraphics[width=0.42\textwidth]{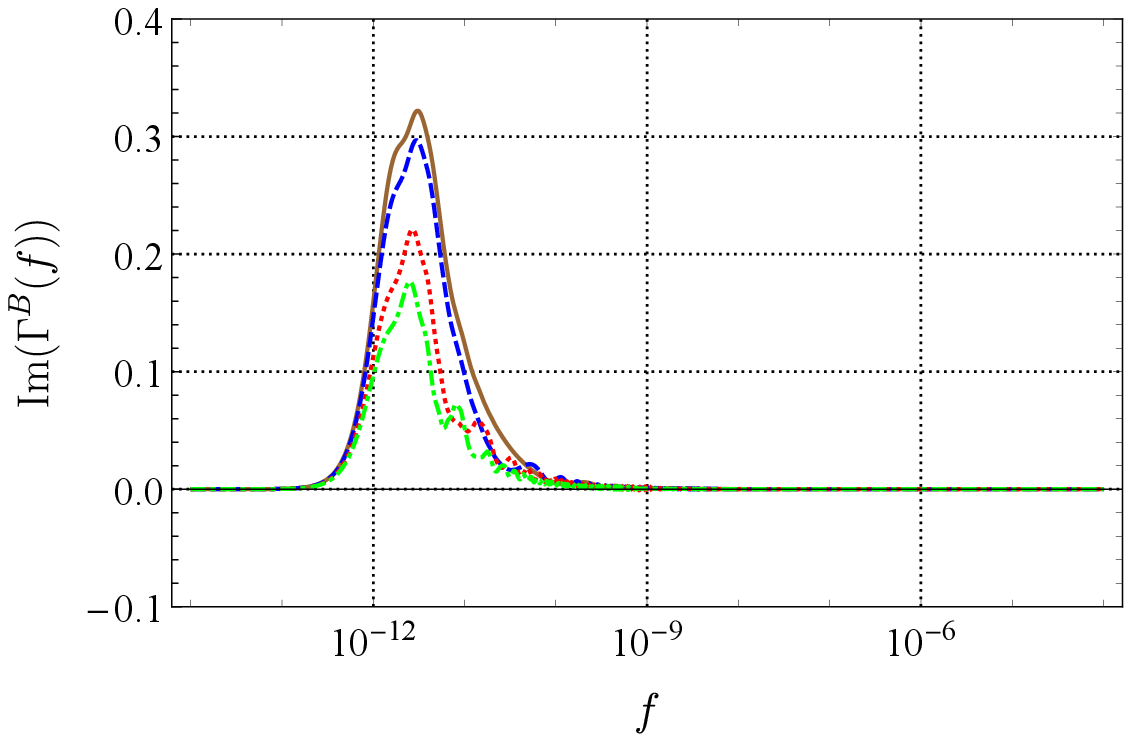} \\
    \includegraphics[width=0.42\textwidth]{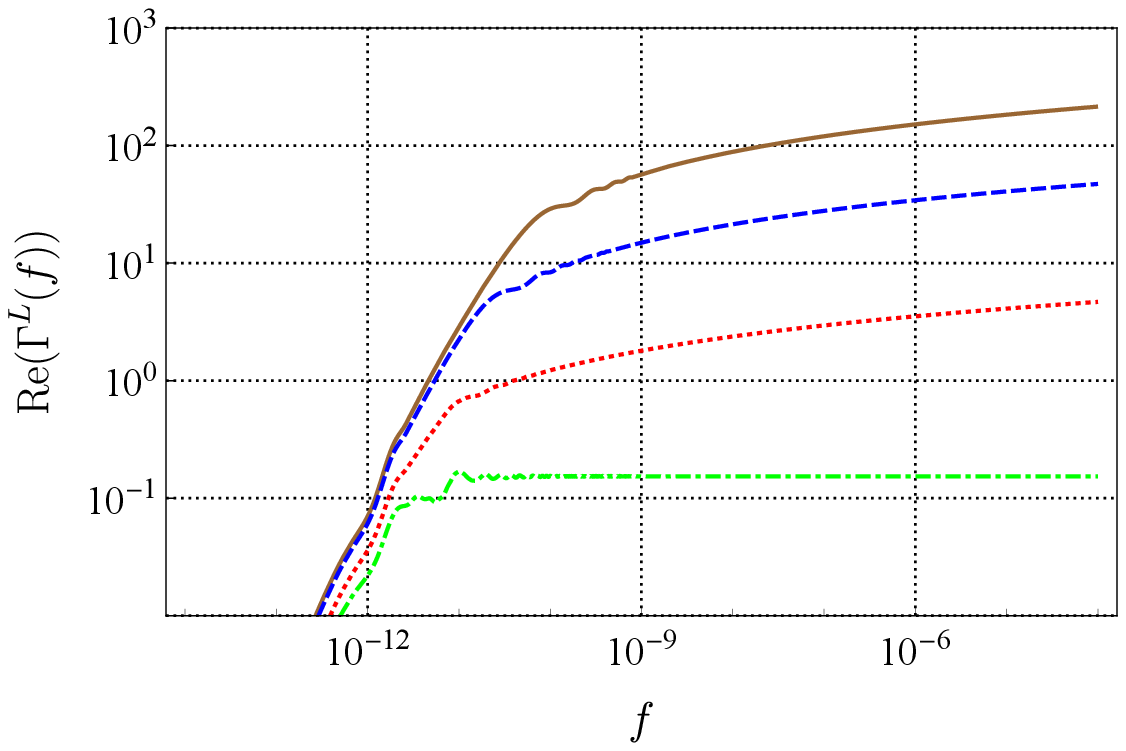}
    \includegraphics[width=0.42\textwidth]{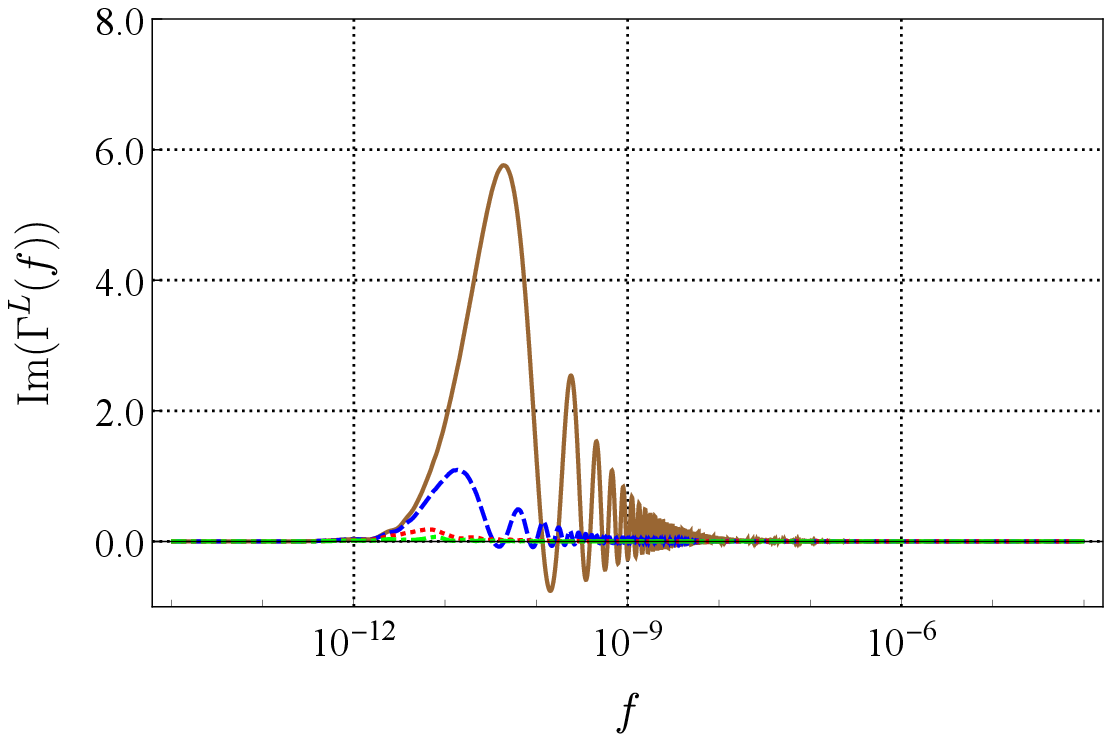} 
    \caption{The GWB frequency dependence of the OFR for the different modes.
    The real and imaginary parts of the ORF for different modes are shown in different picture.
    The tensor mode and scalar-breathing mode are normalized. The distance between the two pulsars and the Earth are fixed as $L_u=1kpc$ and $L_v=4kpc$.
    The different lines represent different angles between pulsars: $\gamma=\pi/12$ (brown solid), $\gamma=\pi/6$ (green dashed),  $\gamma=\pi/3$ (red dotted) and $\gamma=\pi/2$ (green dotdashed).}\label{fig4}
\end{figure*}
Employing the method in \cite{2019prd_analytic_response_function_TDI,2021prd_sensitivity_TDI,2021prd_sensitivity_TDI_nonGR}, we obtained the full analytic expression of the ORF for PTA without any approximation for all possible polarizations allowed by the general metric theory of gravity.
Since no approximation is used, for any two pulsars, the ORF can be obtained quickly and accurately by the expression.
For example, plots of the ORF of a pair of pulsars for different modes are shown in Fig. \ref{fig4}, plotted as functions of frequency $f$.
We choose to fix the distance between the two pulsars and the Earth $L_u=1kpc$ and $L_v=4kpc$, and draw different curves at different angular.
In the current detection band of PTA of $10^{-9} - 10^{-6}$ Hz, the ORF for tensor mode and scalar-breathing mode can be regarded as a constant that only depends on the angular separation $\gamma$ between pairs of pulsars.
However, the ORF for vector mode at small $\gamma$ is likely to vary with frequency.
And for the scalar-longitudinal mode, the ORF grows logarithmically with frequency if $\gamma<\pi/2$.

The tensor mode and scalar-breathing mode show similar behavior, since they are both transverse modes.
And they are quite different from the longitudinal mode (vector and scalar-longitudinal).
For co-directional pulsars ($\gamma=0$), the ORF of transverse modes is finite, no matter how large $fL$ is.
However, the ORF of longitudinal modes grows with $fL$.
The vector mode grows logarithmically with $fL$, and the scalar-longitudinal mode grows linearly.
In addition, the phases of the two kinds of modes are also different.
In the actual detection, the distances from the two pulsars to Earth are usually not the same, and the ORF is complex in this case.
As long as $fL$ is large enough, the imaginary part of the ORF for transverse modes can always be ignored.
For longitudinal modes, this can only be done around $\gamma=\pi/2$.

For most cases, the short-wave expressions \cite{1983apj_HD_curve, 2008apj_Lee_ORF_nonGR} are valid and succinct, and our result will reduce to them with short-wave approximation.
For the scalar-longitudinal mode, there is no short-wave expression except for the co-aligned pulsars $\gamma=0$.
But our results hold for scalar-longitudinal modes at any angle separation.
The auto-correlated response functions, which are inferred from our results, also agree with the prior literature \cite{2008apj_Lee_ORF_nonGR,2012prd_ORF_nonGR} for the co-aligned pulsars.
Compared with the recent series expansions \cite{2021prd_ORF_PTA_series_GR,2022prd_ORF_PTA_series_nonGR},  the numerically calculated values of our expressions are consistent, and our computation time will be much less.
The only thing to worry about is that one can't bring that value directly into the calculation for $\gamma=0$, as that could lead to uncertainty. 
In this case, the expressions under the limit $\gamma \rightarrow 0$ are needed, that is, the auto-correlated response functions.

Totally, for current PTA detection, our results improve slightly for transverse modes.
The short-wave approximation is very precise and concise for most of the pulsars.
Our results only improve for nearly co-located pulsars, such as pulsar binary systems, and pulsars close to Earth that may be discovered in the future.
However, our results are very meaningful for other modes in modified gravity theory, especially the scalar-longitudinal mode.
For vector mode, the short-wave approximation is not valid for small angular separation between pulsars.
And the situation is even worse for scalar-longitudinal mode, where the shortwave approximation is not valid for any angle separation.
Our results greatly improve this situation and provide help for future detection of the polarized GWB.
The full analytic expression of ORF is very useful in GWB data analysis of PTA.

\section*{Acknowledgments}
This work is supported by the National Natural Science Foundation of China (Grant Nos. 11925503 and 12175076 ) and Guangdong Major
project of Basic and Applied Basic Research (Grant No. 2019B030302001).

\appendix*
\section{Detailed calculations for the integration of ORF\label{sec5}}
Through the variable transformation from $\{\theta,\phi\}$ to $\{x,y\}$, the primitive function of the integral of the ORF can be found.
For example, for the integral in the ORF for scalar-breathing mode,
\begin{widetext}
\begin{equation}\label{I}
    \begin{aligned}
    I_1=&\int_{-1}^1dx\int_{y_l}^{y_u}dy \frac{(1-x)(1-y)}{\eta(x,y)} \\
    =&\int_{-1}^1dx (1-x)\left((1-x\cos\gamma)\arctan\left(\frac{y-x\cos\gamma}{\eta(x,y)}\right)+\eta(x,y)\right)\bigg|_{y_l}^{y_u} \\
    =&\pi\int_{-1}^1dx (1-x)(1-x\cos\gamma) \\
    =&2\pi(1+\frac{\cos\gamma}{3}) ,
    \end{aligned}
\end{equation}
where 
$$
\begin{aligned}
    \int \frac{(1-x)(1-y)}{\eta(x,y)} dy
    &=(1-x)\int \left(\frac{1-x\cos\gamma}{\eta(x,y)}-\frac{y-x\cos\gamma}{\eta(x,y)} \right)dy \\
    &=(1-x)\left((1-x\cos\gamma)\arctan\left(\frac{y-x\cos\gamma}{\eta(x,y)}\right)+\eta(x,y)\right) ,
\end{aligned} 
$$
$$\eta(x,y_l)=\eta(x,y_u)=0 \quad
\lim_{y\rightarrow y_l,y_u}\arctan\left(\frac{y-x\cos\gamma}{\eta(x,y)}\right)=\pm \pi/2,$$
are used to obtain the second and third line respectively.
For the exponential terms, the method is also valid,
\begin{equation}\label{I2}
    \begin{aligned}
    I_2=&-\int_{-1}^1dx\int_{y_l}^{y_u}dy e^{-i\beta_u(1+x)}\frac{(1-x)(1-y)}{\eta(x,y)} \\
    =&\int_{-1}^1dx e^{-i\beta_u(1+x)}(x-1)\left((1-x\cos\gamma)\arctan\left(\frac{y-x\cos\gamma}{\eta(x,y)}\right)+\eta(x,y)\right)\bigg|_{y_l}^{y_u} \\
    =&\pi \int_{-1}^1dx e^{-i\beta_u(1+x)}(x-1)(1-x\cos\gamma) \\
    =&\frac{2i\pi(1+\cos\gamma)}{\beta_u}+\frac{\pi\left(e^{-2i\beta_u}(1-\cos\gamma)-1-3\cos\gamma\right)}{\beta_u^2}
    +\frac{2i\pi\cos\gamma(e^{-2i\beta_u}-1)}{\beta_u^3} .
    \end{aligned}
\end{equation}
If the exponential term contains $y$, just change the integral order,
\begin{equation}\label{I3}
    \begin{aligned}
    I_3=&-\int_{-1}^1dx\int_{y_l}^{y_u}dy e^{i\beta_v(1+y)}\frac{(1-x)(1-y)}{\eta(x,y)} \\
    =&-\int_{-1}^1dy\int_{x_l}^{x_u}dx e^{i\beta_v(1+y)}\frac{(1-x)(1-y)}{\eta(x,y)} \\
    =&\int_{-1}^1dy e^{i\beta_v(1+y)}(y-1)\left((1-y\cos\gamma)\arctan\left(\frac{x-y\cos\gamma}{\eta(x,y)}\right)+\eta(x,y)\right)\bigg|_{x_l}^{x_u} \\
    =&\pi\int_{-1}^1dy e^{i\beta_v(1+y)}(y-1)(1-y\cos\gamma) \\
    =&-\frac{2i\pi(1+\cos\gamma)}{\beta_v}+\frac{\pi\left(e^{2i\beta_v}(1-\cos\gamma)-1-3\cos\gamma\right)}{\beta_v^2}
    -\frac{2i\pi\cos\gamma(e^{2i\beta_v}-1)}{\beta_v^3} ,
    \end{aligned}
\end{equation}
where  $x_l,x_u=y\cos\gamma \mp \sin\gamma \sqrt{1-y^2} $.
The same result can be easily obtained by changing $\beta_u$ in (\ref{I2}) to $-\beta_v$, because the integration region $\eta(x,y)\geq 0$ is symmetric for $x$ and $y$.
To calculate the integral that includes both $x$ and $y$ in the exponential term, we change the variables $\{x,y\}$ to 
$a=x\sin\chi-y\cos\chi$ and $b=x\cos\chi+y\sin\chi $, where $\chi=\arctan(\beta_u/\beta_v)$.
And the integration region becomes $\tilde{\eta}(a,b) \geq 0$.
It rotates the $xy$ plane by an angle $\chi$ such that $-x\beta_u+y\beta_v = -\beta (x\sin\chi-y\cos\chi)=-a\beta$, where $\beta=\sqrt{\beta_u^2+\beta_v^2}$.
With the new variables, the integration is
\begin{equation}\label{I4ab}
    \begin{aligned}
    I_4&=\int e^{-i\beta_u(1+x)+i\beta_v (1+y)}\frac{(1-x)(1-y)}{\eta(x,y)}\\
       &=\int_{-\sqrt{1-\cos\gamma\sin2\chi}}^{\sqrt{1-\cos\gamma\sin2\chi}}da \int_{b_d}^{b_u}db
       \frac{e^{-i\beta(a+\sin\chi-\cos\chi)}}{\tilde{\eta}(a,b)}
       (1-a\sin\chi-b\cos\chi)(1+a\cos\chi-b\sin\chi).
    \end{aligned}
\end{equation}
where 
$$\tilde{\eta}(a,b)=\sqrt{\sin^2\gamma-2ab\cos\gamma\cos 2\chi-a^2(1+\cos\gamma\sin 2\chi)-b^2(1-\cos\gamma\sin 2\chi)} ,$$
and $b_d,b_u=\frac{-a\cos\gamma\cos2\chi \mp \sin\gamma\sqrt{1-\cos\gamma\sin2\chi-a^2}}{1-\cos\gamma\sin2\chi}$, which can be obtained by solving $\tilde{\eta}(a,b)=0$.
The rest are similar to the previous process. 
$\tilde{\eta}(a,b)$ can be expressed in the form $l\sqrt{m-(b-n a)^2}$.
In this way the integral for $b$ can be decomposed into three parts: $\int \frac{(b-n a)^2}{\sqrt{m-(b-n a)^2}} db$, 
$\int \frac{b-n a}{\sqrt{m-(b-n a)^2}} db$ and $\int \frac{1}{\sqrt{m-(b-n a)^2}} db$.
The remaining integral with respect to $a$ is a product of an exponential and a polynomial.
After some tedious process, the final result is 
\begin{equation}\label{I4}
    \begin{aligned}
        I_4=&\frac{\pi e^{-i(\beta_u-\beta_v)}}{\beta_u \beta_v} 
        \left[\sin\beta_s\left(\frac{2\beta^2+(\beta_u^2-\beta_v^2)^2}{2\beta_s^3}
        -\frac{3(\beta_u^2-\beta_v^2)^2}{2\beta_s^5}\right. \right.\\
        &\left.+\frac{i(\beta_u-\beta_v)(\beta_u+\beta_v)^2}{\beta_s^3}
        +\frac{1+4\beta_u \beta_v-2i(\beta_u \beta_v)}{2\beta_s}-\frac{\beta_s}{2}\right) \\
        &\left.+\cos\beta_s \left(-\frac{1}{2}-\frac{\beta^2}{\beta_s^2}
        +\frac{3(\beta_u^2-\beta_v^2)^2}{2\beta_s^4}-\frac{i(\beta_u-\beta_v)(\beta_u+\beta_v)^2}{\beta_s^2}+i(\beta_u-\beta_v)\right)\right] .
    \end{aligned}
\end{equation}
So the ORF for scalar-breathing mode is 
\begin{equation}
    \Gamma^B_{IJ}(f)=\frac{1}{8\pi}(I_1+I_2+I_3+I_4) .
\end{equation}

For the integral in the ORF for tensor mode in Eq. \ref{ORF_T_xy}, the first term is the same as the scalar-breathing mode,
and the rest term is 
$$-2\int_{-1}^1dx\int_{y_l}^{y_u}dy 
\frac{(1-e^{-i\beta_u (1+x)})(1-e^{i\beta_v (1+y)})}{(1+x)(1+y)}\eta(x,y). $$
It can be divided into four parts in the previous way.
\begin{equation}\label{I5}
    \begin{aligned}
    I_5=&-2\int_{-1}^1dx \frac{1}{(1+x)} \int_{y_l}^{y_u}dy \frac{\eta(x,y)}{(1+y)}\\
    =&2\int_{-1}^1dx \frac{1}{(1+x)} \int_{y_l}^{y_u}dy \left(\frac{(x+\cos\gamma)^2}{(1+y)\eta(x,y)}
    -\frac{1+x \cos\gamma}{\eta(x,y)}+\frac{y-x\cos\gamma}{\eta(x,y)}\right)\\
    =&2\pi\left(-\int_{-1}^{-\cos\gamma} (1+\cos\gamma)dx +2\sin^2\frac{\gamma}{2}\int_{-\cos\gamma}^{1}\frac{(x-1)}{1+x}dx\right)\\
    =&16\pi\sin^2\frac{\gamma}{2} \ln\sin\frac{\gamma}{2}
    \end{aligned}
\end{equation}
where the integration in second line is
$$ \int \frac{(x+\cos\gamma)^2}{(1+y)\eta(x,y)} dy =(x+\cos\gamma)\arctan\frac{(1-x^2)\sin^2\gamma-(1+x\cos\gamma)(x\cos\gamma-y)}{(x+\cos\gamma)\eta(x,y)} , $$
$$ \int \frac{1+x \cos\gamma}{\eta(x,y)} dy =(1+x\cos\gamma)\arctan\frac{y-x\cos\gamma}{\eta(x,y)} ,$$
$$ \int \frac{y-x\cos\gamma}{\eta(x,y)} dy =-\eta(x,y). $$
\begin{equation}\label{I6}
    \begin{aligned}
    I_6=&2\int_{-1}^1dx \frac{e^{-i\beta_u(1+x)}}{(1+x)} \int_{y_l}^{y_u}dy\frac{\eta(x,y)}{(1+y)}\\
    =&2\pi\left(\int_{-1}^{-\cos\gamma} e^{-i\beta_u(1+x)}(1+\cos\gamma)dx 
    -2\sin^2\frac{\gamma}{2}\int_{-\cos\gamma}^{1}e^{-i\beta_u(1+x)}\frac{(x-1)}{1+x}dx\right)\\
    =&\frac{2i\pi}{\beta_u}\left(2e^{-i\beta_u(1-\cos\gamma)}-(1+\cos\gamma)-(1-\cos\gamma)e^{-2i\beta_u}\right)
    +8\pi \sin^2\frac{\gamma}{2}(\text{Ei}(-2i\beta_u)-\text{Ei}(-i\beta_u(1-\cos\gamma))) .
    \end{aligned}
\end{equation}
\begin{equation}\label{I7}
    \begin{aligned}
    I_7=&2\int_{-1}^1dx \frac{1}{(1+x)} \int_{y_l}^{y_u}dy\frac{e^{i\beta_v(1+y)}\eta(x,y)}{(1+y)}\\
    =&2\int_{-1}^1dy \frac{e^{i\beta_v(1+y)}}{(1+y)} \int_{x_l}^{x_u}dx\frac{\eta(x,y)}{(1+x)}\\
    =&\frac{2i\pi}{\beta_v}\left(1+\cos\gamma-2e^{i\beta_v(1-\cos\gamma)}+(1-\cos\gamma)e^{2i\beta_v}\right)
    +8\pi \sin^2\frac{\gamma}{2}(\text{Ei}(2i\beta_v)-\text{Ei}(i\beta_v(1-\cos\gamma))).
    \end{aligned}
\end{equation}
\begin{equation}
    \begin{aligned}
    I_8=&-2\int_{-1}^1dx\int_{y_l}^{y_u}dy \frac{e^{-i\beta_u(1+x)+i\beta_v(1+y)}\eta(x,y)}{(1+x)(1+y)}\\
    =&-2\int_{-\sqrt{1-\cos\gamma\sin2\chi}}^{\sqrt{1-\cos\gamma\sin2\chi}}da \int_{b_d}^{b_u}db
    \frac{e^{-i\beta(a+\sin\chi-\cos\chi)}\tilde{\eta}(a,b)}{(1+a\sin\chi+b\cos\chi)(1-a\cos\chi+b\sin\chi)}.
    \end{aligned}
\end{equation}
It can be simplified as the form $\int \frac{\sqrt{m-(b-n a)^2}}{p-b} db $ and the rest process is trivial.
The result is 
\begin{equation}
    \begin{aligned}
    I_8=&\frac{4\pi e^{i(\beta_v-\beta_u)}}{\beta_u \beta_v}\left(\beta_s\sin\beta_s-i(\beta_u-\beta_v)\cos\beta_s\right)
    -4i\pi\left(\frac{e^{-i\beta_u(1-\cos\gamma)}}{\beta_u}-\frac{e^{i\beta_v(1-\cos\gamma)}}{\beta_v}\right)\\
    -&8\pi \sin^2\frac{\gamma}{2}\left(\text{Ei}(i(\beta_v-\beta_u-\beta_s))+\text{Ei}(i(\beta_v-\beta_u+\beta_s))
    -\text{Ei}(-i\beta_u(1-\cos\gamma))-\text{Ei}(i\beta_v(1-\cos\gamma))\right) .
    \end{aligned}
\end{equation}
Finally, the ORF of the tensor modulus is obtained,
\begin{equation}
    \Gamma^T_{IJ}(f)=\frac{1}{16\pi}\sum_{i=1}^8 I_i.
\end{equation}

The above method can be utilized to obtain the ORF for all polarizations.
It should be noted that the process for scalar-longitudinal mode is slightly different.
Divide the integral into separate parts, and the separate integration of each part will get divergent results.
We need to drop out the divergent parts in the integral result.
And it can be proved that the divergent parts dropped out just cancel out. 
\end{widetext}

\bibliography{reference}
\bibliographystyle{h-physrev.bst}

\end{document}